\documentclass[twocolumn,english]{revtex4-1}
\usepackage[T1]{fontenc}
\usepackage[latin9]{inputenc}
\usepackage{textcomp}
\usepackage{amsmath}
\usepackage{amssymb}
\usepackage{graphicx}
\usepackage{esint}

\makeatletter

\providecommand{\tabularnewline}{\\}

\makeatother

\usepackage{babel}
\begin{document}

\title{A method to efficiently simulate the thermodynamic properties of
the Fermi-Hubbard model on a quantum computer}

\author{Pierre-Luc Dallaire-Demers}

\author{Frank K. Wilhelm}

\affiliation{Theoretical physics, Saarland University, Saarbrücken, 66123 Germany}

\date{\today}
\begin{abstract}
Many phenomena of strongly correlated materials are encapsulated in
the Fermi-Hubbard model whose thermodynamic properties can be computed
from its grand canonical potential. In general, there is no closed
form expression of the grand canonical potential for lattices of more
than one spatial dimension, but solutions can be numerically approximated
using cluster methods. To model long-range effects such as order parameters,
a powerful method to compute the cluster's Green's function consists
in finding its self-energy through a variational principle. This allows
the possibility of studying various phase transitions at finite temperature
in the Fermi-Hubbard model. However, a classical cluster solver quickly
hits an exponential wall in the memory (or computation time) required
to store the computation variables. Here it is shown theoretically
that the cluster solver can be mapped to a subroutine on a quantum
computer whose quantum memory usage scales linearly with the number
of orbitals in the simulated cluster and the number of measurements
scales quadratically. A quantum computer with a few tens of qubits
could therefore simulate the thermodynamic properties of complex fermionic
lattices inaccessible to classical supercomputers.
\end{abstract}


\maketitle

\section{Introduction\label{sec:Introduction}}

The Fermi-Hubbard model (FHM) \citep{Hubbard63} is a central tool
in the study of strongly correlated electrons in condensed matter
physics \citep{Sachdev11}. It captures the simplest essence of the
atomic structure of materials and the second quantization of the many-body
interacting wavefunction and can be used to model phase transitions
in Mott insulators, high-$T_{c}$ superconductors \citep{Guillot07,Kaczmarczyk13},
heavy-fermion compounds \citep{Masuda15}, atoms in optical lattices,
organic materials and many others. The exact solutions to the one-dimensional
Hubbard model are known and well understood \citep{Voit95,Lieb03,Essler05}
but the two- and three-dimensional models are known not to have general
closed form solutions and are subject to important theoretical studies
\citep{Uglov94,Tasaki98,Senechal00,Senechal04,Kurzyk07}. An elegant
approximation method valid for short-range interactions is cluster
perturbation theory (CPT), where a lattice is divided into manageable
identical clusters which are solved and then recomposed into a lattice
through with perturbation theory \citep{Senechal00,Senechal08}. However,
the method is not sufficient to systematically account for broken
symmetries in the FHM and has to be extended. In superconductors and
antiferromagnets, local interactions can have long-range effects and
order parameters can appear in different regions of phase space. These
effects can be taken into account in the Green's function of a cluster
by finding the stationary point of the lattice's grand canonical potential
when the self-energy of a cluster is taken as the variational parameter
\citep{Potthoff06}. This self-energy functional theory (SFT) is a
great computational tool to study the important macroscopic thermodynamic
phases of the Hubbard model starting from its microscopic description.
In the context of the SFT, the CPT approximation is generalized to
what is known as the variational cluster approximation (VCA).

However even simulating a small cluster with a handful of electrons
(or orbitals) is a difficult task for classical computers since the
matrices involved in the computation scale exponentially in size with
respect to the number of electronic orbitals. The quantity of information
involved in the precise numerical treatment of large strongly correlated
electronic systems quickly reaches magnitudes where no reasonnable
classical memory technology is sufficient to store it all. Therefore,
being given access to a large controllable Hilbert space in a quantum
computer offers the possibility of simulating electronic systems at
the microscopic level with a greater complexity and accuracy than
the ones accessible to classical computers \citep{Feynman82}. 

This work is inspired from recently developed approaches in quantum
simulations such as the simulation of spin systems \citep{LasHeras13,Salathe15},
fermionic systems and quantum chemistry \citep{Peruzzo14,LasHeras15,Barends15}
and boson sampling to extract vibronic spectra \citep{Huh14}. In
general, it happens that the occupation state of an electronic orbital
can be efficiently represented by one qubit on a quantum computer
through the Jordan-Wigner transformation. The memory bottleneck in
numerically representing the many-body wavefunction is overcome by
making sure that it is never measured and stored on a classical memory
at any point during the simulation. In the VCA, the quantities that
need to be extracted from the wavefunction are the intra-cluster single-particle
correlation functions whose number scales quadratically with the number
of orbitals in a given cluster. On the practical side, it is not yet
known how the computing power of quantum processing devices will scale
in the future, but machines with a fews tens or hundreds of qubits
could already be very useful to run quantum subroutines as part of
larger classical simulation algorithms. This proposed method could
open a practical way to model and engineer the electronic behavior
of strongly-correlated materials with intricate crystalline structures
in a unified and consistent manner. Furthermore the underlying SFT
is very general \citep{Tong05,Filor14} and not restricted to the
class of FHMs. Similar schemes to simulate spin systems, the Bose-Hubbard
model or more exotic fields in lattice gauge theories \citep{Zohar15,Zohar15b}
can likely be constructed in a similar fashion.

This paper aims at at being self-containend by providing all the concepts
required to implement the solver on a general purpose quantum computer
\citep{DiVincenzo2000}. It is structured in the following manner.
Section \ref{sec:VariationalClusterApproximation} summarizes the
variational cluster method used to compute properties of the FHM.
In subsection \ref{sub:GrandCanonicalPotentialFunctionalSelfEnergy},
a variational principle of the self-energy for the grand canonical
potential of the model is outlined such that it can account for possible
long-range ordering effects. Subsection \ref{sub:VariationalCluster}
formalizes the approximation where the Fermi-Hubbard lattice is divided
in independent clusters linked with hopping terms. Section \ref{sec:ExampleSuperconductivity}
introduces the detailed formal description of a cluster using the
example of a 2D lattice with superconductivity starting in subsection
\ref{sub:HamiltonianCluster}. Subsection \ref{sub:SuperlatticeClusters}
proceeds with reviewing the formalism to compute the Green's function
of the lattice from the independent clusters and subsection \ref{sub:CalculationOfObservables}
lists methods to compute observables of interest once the variational
problem is solved. Section \ref{sec:EigenvalueProblem} covers the
computer intensive step where the eigenvalue problem of the cluster
Hamiltonian must be solved at each iteration of the variational solver.
Subsection \ref{sub:ClassicalComputer} summarizes the solution method
on a classical computer and a memory efficient quantum subroutine
to introduced in subsection \ref{sub:QuantumComputer}. The procedure
to measure the Green's function of the cluster is described in subsection
\ref{sub:MeasuringCorrelationFunction}. Appendix \ref{sec:NumericalExample}
presents numerical results where the quantum procedure to compute
a cluster's Green's function is shown to be equivalent to traditional
solution methods. In appendix \ref{sec:GibbsStatePreparation}, details
of the initial Gibbs state preparation are given for a specific algorithm.

\section{Solving the Fermi-Hubbard model with the variational cluster approximation\label{sec:VariationalClusterApproximation}}

The goal of this section is to introduce the important physical quantities
of the main loop of the numerical variational solver used to extract
properties of the FHM. Since the interesting observables typically
correspond to the response of the system to external perturbations,
the central object of study is the Green's function which contains
both the thermal and the dynamical properties of the system. To compute
the Green's function, a variational principle on the grand canonical
potential is derived from functional arguments. The Green's functional
variational problem is then mapped to a self-energy variational problem
to account for possible spontaneous symmetry breaking from long-range
ordering in a self-consistent manner. At last the lattice approximation
is introduced to complete the description of the lattice variational
solver.

\subsection{The grand canonical potential as a functional of the self-energy\label{sub:GrandCanonicalPotentialFunctionalSelfEnergy}}

Variational solvers \citep{Senechal08} are powerful tools to solve
many-body problems in quantum mechanics. The FHM is an effective description
of the microscopic physics of the electrons in a solid useful in calculating
the properties of Fermi liquids, Mott insulators, anti-ferromagnets
\citep{Rickayzen91}, superconductors \citep{Leggett06} and other
metallic phases. The model describes a simple electronic band in a
periodic lattice $\Gamma$ where electrons are free to hop between
orbitals (or sites) with kinetic energy $t$ and interact via a simple
two-body Coulomb term $U$. The standard form of the Fermi-Hubbard
Hamiltonian is given by

\begin{equation}
\mathcal{H}=-t\sum_{\left\langle i,j\right\rangle ,\sigma}c_{i\sigma}^{\dagger}c_{j\sigma}-U\sum_{i}n_{i\uparrow}n_{i\downarrow}-\mu\sum_{i,\sigma}n_{i\sigma},\label{eq:FermiHubbardHamiltonian}
\end{equation}
where $\mu$ is the chemical potential that determines the occupation
of the band. The $c_{i\sigma}$($c_{i\sigma}^{\dagger}$) are the
fermionic annihilation (creation) operators and the number operators
are $n_{i\sigma}=c_{i\sigma}^{\dagger}c_{i\sigma}$. Note that in
the rest of this document, units are used such that $t\rightarrow1$
is assumed to be the reference energy and inverse time. It is also
assumed that $\hbar\rightarrow1$ and $k_{B}\rightarrow1$.

\subsubsection{The Luttinger-Ward formalism\label{sub:LuttingerWardFormalism}}

\begin{figure}
\begin{centering}
\includegraphics[width=3.375in]{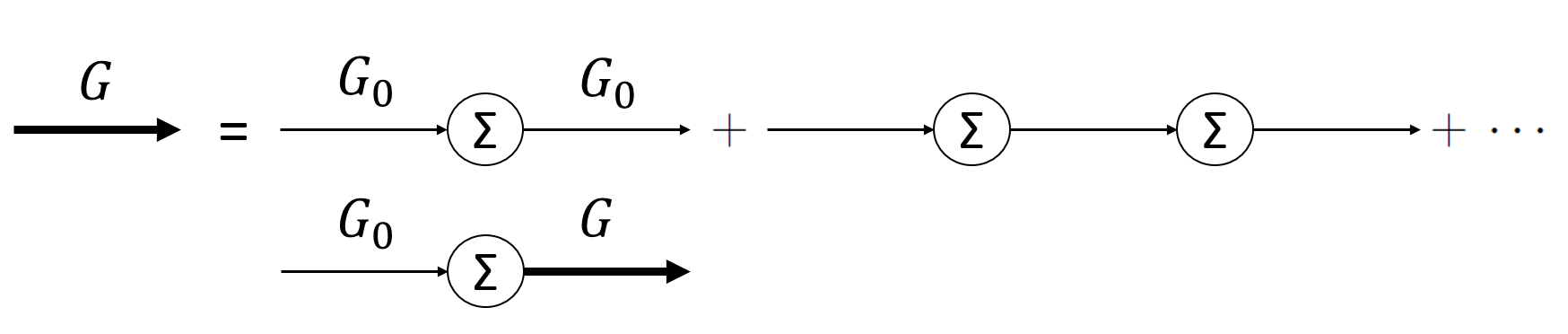}
\par\end{centering}

\caption{Diagrammatic representation of the relation between the single-particle
Green's function $\mathbf{G}$, the bare Green's function of the non-interacting
lattice $\mathbf{G_{0t}}$ and the self-energy $\mathbf{\Sigma}$.\label{fig:GreensFunction}}
\end{figure}
The Green's function $\mathbf{G}$ of the full system described by
$\mathcal{H}$ can be obtained exactly from the bare single-particle
Green's function of the non-interacting lattice (tight-binding) $\mathbf{G_{0t}}$
and the self-energy $\mathbf{\Sigma}=\mathbf{G_{0t}^{-1}}-\mathbf{G^{-1}}$
by solving the Dyson equation represented in figure \ref{fig:GreensFunction}

\begin{equation}
\mathbf{G}=\mathbf{G_{0t}}+\mathbf{G_{0t}\mathbf{\Sigma}\mathbf{G}}.\label{eq:DysonEquationFullLattice}
\end{equation}
When there is no interaction, the self-energy is zero and the tight-binding
Green's function for a given one-body hopping matrix $\mathbf{t}$
is
\begin{equation}
\mathbf{G_{0t}^{-1}}=\omega-\mathbf{t}.\label{eq:TightBindingGreensFunction}
\end{equation}

\begin{figure}
\begin{centering}
\includegraphics[width=3in]{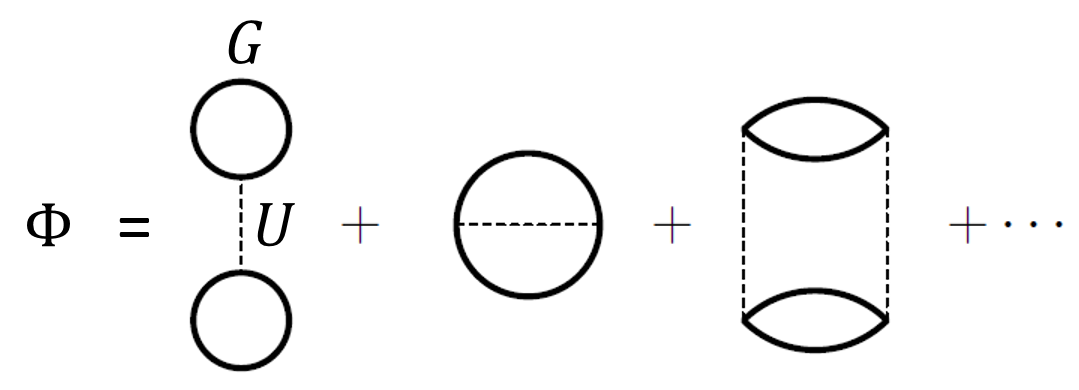}
\par\end{centering}

\caption{The Luttinger-Ward functional $\Phi$ is the sum of all the two-body
skeleton diagrams. The functional derivative with respect to $\mathbf{G}$
gives all the diagrams for the computation of $\mathbf{\Sigma}$.
In the case where $U=0$ then $\Phi\left[\mathbf{G}\right]=0$. \label{fig:LuttingerWardFunctional}}

\end{figure}

The model can be considered ``solved'' once the single-particle
Green's function $\mathbf{G}$ can be computed accurately for any
interesting input coordinates (such as position / momentum, time /
energy). A method to obtain the Green's function consists in rewriting
the Dyson equation as a variational principle on the grand canonical
potential of the system. To accomplish this task, it is useful to
introduce the Luttinger-Ward functional \citep{Potthoff06,RiosHuguet07}
of the Green's function $\Phi\left[\mathbf{G}\right]$ which generates
all two-body skeleton diagrams (see figure \ref{fig:LuttingerWardFunctional})
and has the interesting property that its functional derivative with
respect to $\mathbf{G}$ is simply

\begin{equation}
\frac{\delta\Phi\left[\mathbf{G}\right]}{\delta\mathbf{G}}=\mathbf{\Sigma}.\label{eq:LuttingerWardFunctional}
\end{equation}
Furthermore, the functional form of $\Phi\left[\mathbf{G}\right]$
depends only on the form of the interaction $U$ and is independent
of the one-body terms in $\mathcal{H}$. In statistical mechanics,
observables are derived from a thermodynamic potential. For many-body
systems, it is typically easier to let the total number of particles
fluctuate and work with the grand canonical ensemble. The grand canonical
potential of the full lattice can be defined from the Luttinger-Ward
functional as a functional of $\mathbf{G}$

\begin{equation}
\Omega_{t}\left[\mathbf{G}\right]=\Phi\left[\mathbf{G}\right]-\mathrm{Tr}\left[\left(\mathbf{G_{0t}^{-1}}-\mathbf{G^{-1}}\right)\mathbf{G}\right]+\mathrm{Tr}\ln\left[\mathbf{-G}\right],\label{eq:GrandPotentialGreensFunction}
\end{equation}
such that the Dyson equation (\ref{eq:DysonEquationFullLattice})
can be recovered as the stationary point with respect to the variation
of $\mathbf{G}$:

\begin{equation}
\frac{\delta\Omega_{t}\left[\mathbf{G}\right]}{\delta\mathbf{G}}=\mathbf{\Sigma}-\mathbf{G_{0t}^{-1}}+\mathbf{G^{-1}}=0.\label{eq:DysonEqnFromGreensFunction}
\end{equation}

In Ref. \citep{Potthoff05}, Potthoff describes three types of approximation
stategy to solve this variational problem. A type I approximation
would try to simplify the Euler equation from a heuristic argument
but could suffer from thermodynamic inconsistencies. A type II approximation
would correspond to computing the $\Phi\left[\mathbf{G}\right]$ functional
only for a finite set of diagrams, but justifying the use of a particular
functional form over other possibilities is in itself not trivial.
Finally, in a type III approximation, thermodynamical consistency
is preserved as well as the exact form of the Luttinger-Ward functional
but the trial Green's functions are chosen from a restricted domain
where the self-energy is constrained. The VCA is a type III approximation.
The main advantage of this type of scheme is that it allows for a
systematic construction of increasingly accurate solutions to many-body
problems with local interactions. In the case of the FHM, a good scheme
to systematically approximate the self-energy is to consider a reference
lattice of isolated clusters $\Gamma'$ with the same local interaction
term $U$ as the lattice $\Gamma$ and pick $\mathbf{\Sigma}$ from
the exact solution of the reference lattice. This method allows for
the construction of solutions to the FHM that are very accurate except
for long range correlations that exceed the dimensions of the clusters.
The main advantage of this scheme is that the solutions are guaranteed
to become asymptotically exact as the size of the cluster reaches
the size of the original lattice. The next step consists in rewritting
the grand canonical potential $\Omega_{t}$ as a functional of the
lattice self-energy $\Sigma$ instead of the Green's function $\mathbf{G}$.

\subsubsection{Self-energy functional theory \label{sub:SelfEnergyFunctionalTheory}}

The variational principle of the self-energy of a cluster \citep{Potthoff03}
intends to account for solutions of the Hubbard model with spontaneous
symmetry breaking caused by long-range interactions. The grand canonical
potential $\Omega_{t}\left[\mathbf{G}\right]$ can be rewritten as
a functional of the self-energy $\Omega_{t}\left[\mathbf{\Sigma}\right]$
by applying the Legendre transformation $\mathbf{G\left[\mathbf{\Sigma}\right]=\left(\mathbf{G_{0t}^{-1}}-\mathbf{\Sigma}\right)^{-1}}$
such that

\begin{equation}
\begin{array}{ccl}
\Omega_{t}\left[\mathbf{G}\right] & = & \Phi\left[\mathbf{G}\right]-\mathrm{Tr}\left[\left(\mathbf{G_{0t}^{-1}}-\mathbf{G^{-1}}\right)\mathbf{G}\right]+\mathrm{Tr}\ln\left[\mathbf{-G}\right]\\
\\
 & = & \underset{\Lambda\left[\mathbf{\Sigma}\right]}{\underbrace{\Phi\left[\mathbf{G}\right]-\mathrm{Tr}\left[\Sigma\mathbf{G}\right]}}+\mathrm{Tr}\ln\left[\mathbf{-G}\right]\\
\\
 & = & \Lambda\left[\mathbf{\Sigma}\right]-\mathrm{Tr}\ln\left[-\mathbf{G_{0t}^{-1}}+\mathbf{\Sigma}\right]\\
\\
 & = & \Omega_{t}\left[\mathbf{\Sigma}\right].
\end{array}\label{eq:LegendreTransformationGrandPotential}
\end{equation}
Let's then notice that $\Omega_{t}\left[\mathbf{\Sigma}\right]$ is
still exact and now only depends on the self-energy $\Sigma$ and
the non-interacting Green's function $\mathbf{G_{0t}}$. The Legendre
transformed Luttinger-Ward functional $\Lambda\left[\mathbf{\Sigma}\right]$
has the nice property

\begin{equation}
\frac{\delta\Lambda\left[\mathbf{\Sigma}\right]}{\delta\mathbf{\Sigma}}=-\mathbf{G},\label{eq:LambdaSelfEnergyFunctionalDerivative}
\end{equation}
which is used to recover the Dyson equation of the system and the
variational principle depending on the self-energy

\begin{equation}
\frac{\delta\Omega_{t}\left[\mathbf{\Sigma}\right]}{\delta\Sigma}=\left(\mathbf{G_{0t}^{-1}}-\Sigma\right)^{-1}-\mathbf{G}=0.\label{eq:DysonEqnFromSelfEnergy}
\end{equation}
Solutions to the FHM can be found by varying the self-energy until
a physical value of the Green's function is found and the Dyson equation
is satisfied. However, since this is in general a saddle-point problem,
the optimal point cannot be interpreted as an upper bound to the exact
energy (as in the Ritz variational method) but as the most ``physical''
approximation of the grand canonical potential allowed by a given
parametrization of the self-energy. Computing the exact single-particle
self-energy for a large lattice and storing the result are tasks beyond
the capabilities of classical computers. The idea of cluster methods
used to approximate the solution of the full lattice $\Gamma$ is
to divide it into a reference lattice $\Gamma'$ of clusters of a
small number (i.e. computer tractable) of sites, solve a cluster exactly
and use perturbation theory to approximate the properties of the full
lattice.

\subsection{Variational cluster approximation \label{sub:VariationalCluster}}

Large lattices with millions of orbitals are impossible to simulate
exactly on classical computers since the memory required to store
for the associated state vectors scales exponentially in cluster size.
A method to mitigate this problem makes use of the translation invariance
of the lattice. It consists in breaking down the lattice in several
independent clusters and making use of the universality of the Luttinger-Ward
functional to recast the variational equation (\ref{eq:DysonEqnFromSelfEnergy})
on a cluster-restricted domain of the self-energy. The exact solutions
are recovered when the size of the cluster is equal to the size of
the original lattice \citep{Aichhorn06}.

\begin{figure}
\begin{centering}
\includegraphics[width=3.375in]{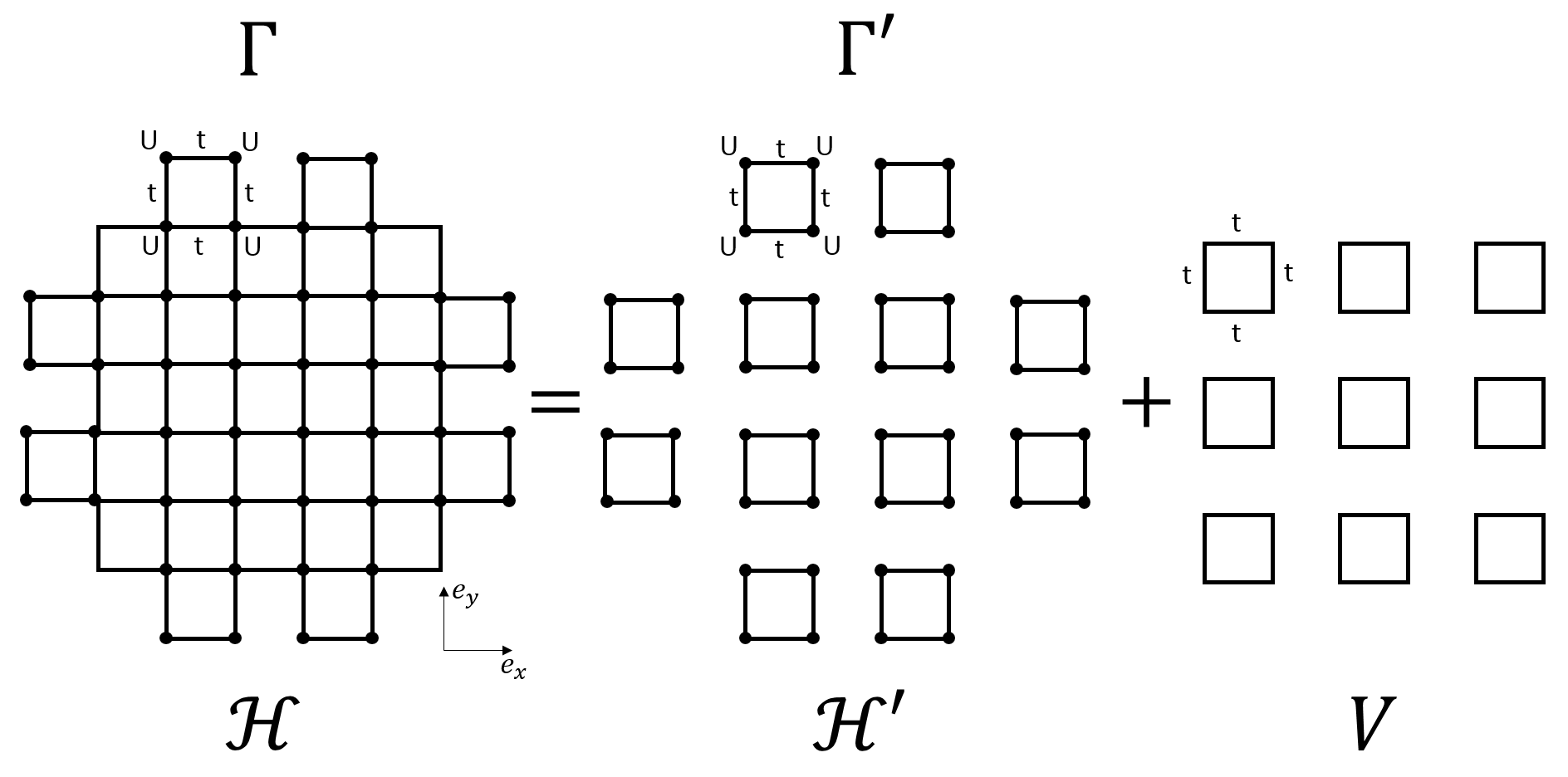}
\par\end{centering}

\caption{The essence of the VCA method is to remove the one-body links (denoted
$t$) between small clusters (contained in $\mathbf{V}$) from the
lattice $\Gamma$ and consider only the reference lattice $\Gamma'$
whose Hamiltonian $\mathcal{H}'$ is block diagonal in the Wannier
basis and easier to solve than the complete problem $\mathcal{H}$.
The reference system generates a manifold of trial self-energies $\Sigma'$
parametrized by single-particle parameters $\mathbf{t'}$. The self-energy
functional can be evaluated exactly on this manifold as the interaction
part of the Hamiltonian (the $U$s) is left unchanged.. The solution
become asymptotically exact as the clusters are made to include more
sites. \label{fig:ClusterPerturbation}}
\end{figure}

Good and thorough introductions to the VCA method can be found in
\citep{Maier05,Senechal08}. In the restricted Hilbert space of a
cluster, the goal is to variationally find a self-energy $\Sigma'$
such that it is most physical (by satisfying the VCA version of the
Dyson equation) and minimizes the free energy. As hinted at the end
of subsection \ref{sub:GrandCanonicalPotentialFunctionalSelfEnergy}
and shown in figure \ref{fig:ClusterPerturbation}, the VCA approximation
consists in subdividing a full lattice $\Gamma$ into a reference
lattice of identical clusters $\Gamma'$ and solving the reference
model exactly in order to obtain its self-energy $\Sigma'$. In this
context, the Green's function of a cluster is a frequency dependent
matrix given by
\begin{equation}
\mathbf{G'^{-1}}\left(\omega\right)=\omega-\mathbf{t'}-\Sigma'\left(\omega\right)\label{eq:ClusterGreensFunction}
\end{equation}
 The Legendre transformed Luttinger-Ward functional $\Lambda$ only
depends on the interaction part of the Hamiltonian. Since by definition
the interaction part of the Hamiltonian is the same for the full system
and the reference system, the identity $\Lambda\left[\mathbf{\Sigma'}\right]=\Lambda\left[\mathbf{\Sigma}\right]$
must hold. Let's note that this scheme would not work directly in
the case of the extended FHM (where there is intersite interaction),
since a reference system of independent clusters cannot be found by
simply removing one-body links of the Hamiltonian \citep{Tong05}.
As in equation (\ref{eq:LegendreTransformationGrandPotential}), the
grand canonical potential of the reference system is given by

\begin{equation}
\Omega'\equiv\Omega_{t'}\left[\mathbf{\Sigma'}\right]=\Lambda\left[\mathbf{\Sigma}\right]-\mathrm{Tr}\ln\left[-\mathbf{G'}\right],\label{eq:GrandPotentialReference}
\end{equation}
where $\mathbf{G'}$ is the Green's function of the reference system.
When they are both evaluated at the self-energy of the reference system,
the difference between the grand canonical potential of the full lattice
and the reference system is

\begin{equation}
\Omega_{t}\left[\mathbf{\Sigma'}\right]=\Omega'+\mathrm{Tr}\ln\left[-\mathbf{G'}\right]-\mathrm{Tr}\ln\left[-\mathbf{G}\right].\label{eq:GrandPotentialDifference}
\end{equation}
 This relation is exact, the only approximation of the VCA is in the
restriction of the domain of the self-energy. It can be further simplified
as the VCA is built within SFT as a well-defined variational extension
to the CPT. The full lattice Green's function $\mathbf{G}\left[\mathbf{\Sigma}\right]$
is equal to the CPT Green's function if its self-energy is restricted
to the domain of the reference system. As in figure \ref{fig:ClusterPerturbation},
it is useful to define $\mathbf{V\equiv\mathbf{t}-}\mathbf{t'}$ as
a perturbation, where $\mathbf{t}$ contains all the one-body terms
of the full lattice $\Gamma$ and $\mathbf{t'}$ represents all the
one-body terms of the lattice of clusters $\Gamma'$. As a result
of strong-coupling perturbation theory, the CPT Green's function is
given by
\begin{equation}
\mathbf{G}\left[\mathbf{\Sigma'}\right]=\mathbf{G_{\mathrm{cpt}}=\left(\mathbf{G'^{-1}}-\mathbf{V}\right)^{-1}}.\label{eq:ClusterPerturbationApproximation}
\end{equation}
With some algebra, equation (\ref{eq:GrandPotentialDifference}) can
be written as

\begin{equation}
\Omega_{t}\left[\mathbf{\Sigma'}\right]=\Omega'-\mathrm{Tr}\ln\left[\mathbf{1}-\mathbf{V}\mathbf{G'}\right].\label{eq:GrandPotentialCPT}
\end{equation}
 The functional is exact as no classes of diagrams have been explicitly
excluded. At the saddle-point, it represents the quantity which is
physically the closest to the physical grand canonical potential of
the full lattice when the self-energy is computed on the reference
lattice. The effect of single-particle correlations and intra-cluster
two-particle correlations is treated non-perturbatively but the inter-cluster
two-particle effects are neglected in the one-particle spectrum. Even
if only a small cluster is exactly solved, the self-energy variational
principle (\ref{eq:DysonEqnFromSelfEnergy}) can be used to study
the properties of the infinite system like the various order parameters
in a thermodynamically consistent framework. Since the VCA is a well
defined generalization of the CPT, it also shares similar characteristics.
It is exact in the limit $\frac{U}{t}\rightarrow0$ where the
self-energy disappears to yield the tight binding model. It is also
exact in the strong-coupling limit $\frac{t}{U}\rightarrow0$,
where all sites are effectively decoupled. The method is easy to generalize
to non-homogenous lattices. The next section introduces the details
of the objects required to compute (\ref{eq:GrandPotentialCPT}) and
find its stationary point as well as some observable that can then
be calculated.

\section{Example on a square lattice with superconductivity \label{sec:ExampleSuperconductivity}}

In this section the self-energy variational approach is used to model
superconductivity in a Fermi-Hubbard lattice. A more general formulation
of possible orders could be made (for arbitrary ordering potentials
and cluster graph), but the goal of this section is only to introduce
the types of formal elements required to describe a cluster. Other
types of order parameters can be found in the literature \citep{Masuda15}.
First the different terms in the Hamiltonian of the cluster are explained.
Then the detailed formalism of the VCA is given through the example
of a square lattice with superconductivity. Finally, various quantities
involved in the computation of useful observables are listed.

\subsection{Hamiltonian of a cluster \label{sub:HamiltonianCluster}}

Each cluster includes only a small portion of the terms of the original
lattice and variational terms must also be included to account for
possible long-range order. For convenience, let's assume that $\Gamma$
is a square lattice with constant spacing $a$. It is broken down
into $N_{c}$ clusters each with $L_{c}$ orbitals (``sites'') with
two electrons each (spin up $\uparrow$ and spin down $\downarrow$).
The Hamiltonian of each cluster is given by

\begin{equation}
\mathcal{H}'=\mathcal{H}_{\mathrm{FH}}+\mathcal{H}_{\mathrm{local}}+\mathcal{H}_{\mathrm{s-pair}}+\mathcal{H}_{\mathrm{d_{x^{2}-y^{2}}}}+\mathcal{H}_{\mathrm{AF}},\label{eq:HamiltonianOneCluster}
\end{equation}
where the Fermi-Hubbard terms remaining in $\Gamma'$ are given by

\begin{equation}
\mathcal{H_{\mathrm{FH}}}=-t\sum_{\left\langle i,j\right\rangle ,\sigma}c_{i\sigma}^{\dagger}c_{j\sigma}-U\sum_{i}n_{i\uparrow}n_{i\downarrow},\label{eq:FermiHubbardHamiltonianCluster}
\end{equation}
which is the same as (\ref{eq:FermiHubbardHamiltonian}) without the
chemical potential term. The chemical potential must be kept as a
variational term to enforce the thermodynamic consistency of the electronic
occupation value

\begin{equation}
\mathcal{H}_{\mathrm{local}}=-\mu'\sum_{i,\sigma}n_{i\sigma}.\label{eq:VariationChemicalPotential}
\end{equation}
It can be seen that at the stationary point $\frac{\partial\Omega_{t}}{\partial\mathbf{t'}}=0$,
the electronic occupation expectation value is 

\begin{equation}
\left\langle n\right\rangle =\mathrm{Tr}\mathbf{\,G}=-\frac{d\Omega_{t}}{d\mu}=-\left(\frac{\partial\Omega_{t}}{\partial\mu}+\frac{\partial\Omega_{t}}{\partial\mathbf{t'}}\cdot\frac{d\mathbf{t'}}{d\mu}\right)\label{eq:ThermodynamicalConsistencyOccupationNumber}
\end{equation}
where the two methods converge to the same average occupation at the
stationary point. Keeping the chemical potential fixed in the cluster
Hamiltonian would break this condition.

The spontaneous transitions of the FHM can be studied by introducing
artificial symmetry breaking terms to the cluster Hamiltonian and
treating them as variational variable. The choice of these terms is
somewhat arbitrary and is usually justified by the physics of the
system studied. For example in the FHM, it is often interesting to
study the competition between superconducting order parameters with
different symmetries and the antiferromagnetic ordering. A variational
singlet pairing term is introduced as

\begin{equation}
\mathcal{H}_{\mathrm{s-pair}}=\Delta'\sum_{i}\left(c_{i\uparrow}^{\dagger}c_{i\downarrow}^{\dagger}+c_{i\downarrow}c_{i\uparrow}\right),\label{eq:sPairingHamiltonian}
\end{equation}
while a $d_{x^{2}-y^{2}}$ singlet pairing takes the form \citep{Senechal08}
\begin{equation}
\mathcal{H}_{\mathrm{d_{x^{2}-y^{2}}}}=\Delta_{d}'\sum_{ij}d_{ij}\left(c_{i\uparrow}^{\dagger}c_{j\downarrow}^{\dagger}+c_{j\downarrow}c_{i\uparrow}\right),\label{eq:dParingHamiltonian}
\end{equation}
where $\mathbf{R}$ are the vector positions of the sites in the cluster
and
\begin{equation}
d_{ij}=\begin{cases}
1 & \mathrm{if}\:\mathbf{R}_{i}-\mathbf{R}_{j}=\pm a\mathbf{e_{x}}\\
-1 & \mathrm{if}\:\mathbf{R}_{i}-\mathbf{R}_{j}=\pm a\mathbf{e_{y}}\\
0 & \mathrm{otherwise.}
\end{cases}
\end{equation}
 The variational Néel antiferromagnetic Weiss field takes the form
\begin{equation}
\mathcal{H}_{\mathrm{AF}}=M'\sum_{i}e^{i\mathbf{Q}\cdot\mathbf{R}_{i}}\left(n_{i\uparrow}-n_{i\downarrow}\right),\label{eq:VariationalAF}
\end{equation}
where $\mathbf{Q}=\left(\pi,\pi\right)$ is the antiferromagnetic
wavevector.

The small parameter in the approximation is $L_{c}^{-1}$, which
means that increasing the size of the cluster also increases the accuracy
of the simulation. \emph{ }

\subsection{The superlattice of clusters \label{sub:SuperlatticeClusters}}

The relation between the original lattice and the lattice of cluster
is given in more details along with useful notations. The main objects
of interest for the quantum subroutine are introduced in this subsection.

\begin{figure}
\begin{centering}
\includegraphics[width=2.5in]{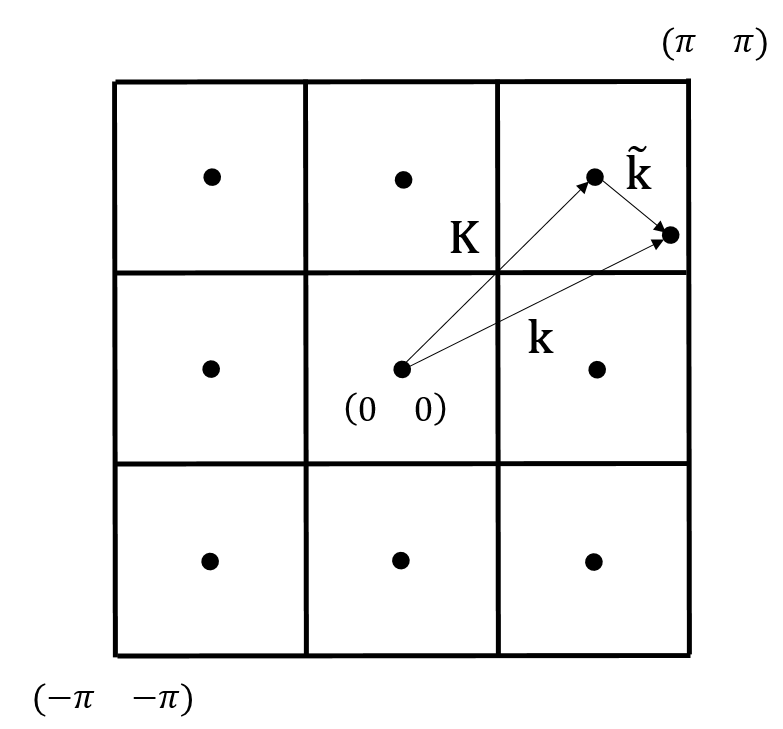}
\par\end{centering}

\caption{Reduced Brillouin zone of the reciprocal lattice. The quasi-momentum
vector $\mathbf{k}$ belong to the reciprocal lattice of $\Gamma$
while the $\mathrm{K}$ component belongs to the reciprocal lattice
of a single cluster. The $\mathbf{\tilde{k}}$ vector belongs to the
reciprocal superlattice (hopping between clusters).\label{fig:ReducedBrilloinZone} }

\end{figure}

\subsubsection{The superlattice in reciprocal space}

To make the procedure clear and concrete, let's work on the example
of the superconducting order parameter on a 2D lattice. For a good
explanation of quantum cluster theories and the details for computations
on clusters of arbitrary size see \citep{Maier05,Senechal08}. A square
lattice with 8 orbitals per cluster is required to study s-wave and
d-wave superconductivity in the FHM. Let's take a lattice $\Gamma$
with $N$ sites and divide it in clusters of $L_{c}=2\times2=4$ sites,
then the number of clusters is simply $N_{c}=\frac{N}{L_{c}}$. Let's
label these 4 sites as $1$, $2$, $3$ and $4$. \emph{} When the
full lattice is Fourier transformed, the first Brillouin zone in quasi-momentum
space is given by

\begin{equation}
\begin{array}{cc}
k_{x/y}=\frac{2\pi m_{x/y}}{Na}, & m_{x/y}=0,\:\ldots,\:N-1\end{array}\label{eq:ReciprocalLattice}
\end{equation}
and the reciprocal superlattice is given by

\begin{equation}
\begin{array}{cc}
\tilde{k}_{x/y}=\frac{2\pi q_{x/y}}{Na}, & q_{x/y}=0,\:\ldots,\:N_{c}-1\end{array}.\label{eq:ReciprocalSuperlattice}
\end{equation}

\subsubsection{The saddle-point problem}

The observable properties of the Hamiltonian (\ref{eq:FermiHubbardHamiltonian})
can be computed from the CPT formula (\ref{eq:ClusterPerturbationApproximation})
by finding variational parameters ($\mu'$,$\Delta'$, $\Delta_{d}'$,
$M'$, etc.) that generate $\Sigma$ for which the Dyson equation
(\ref{eq:DysonEqnFromSelfEnergy}) is stationary. In practice, this
condition is reformulated explicitly over the variational parameters
as
\begin{equation}
\frac{\partial\Omega_{t}}{\partial\mathbf{t'}}=0.\label{eq:GeneralStationarityCondition}
\end{equation}

In the superconducting Fermi-Hubbard example, this would correspond
to solving the saddle-point problem

\begin{equation}
\left(\begin{array}{c}
\frac{\partial\Omega_{t}}{\partial\mathbf{\mu'}}\\
\\
\frac{\partial\Omega_{t}}{\partial\Delta'}\\
\\
\frac{\partial\Omega_{t}}{\partial\Delta_{d}'}\\
\\
\frac{\partial\Omega_{t}}{\partial M'}
\end{array}\right)=\left(\begin{array}{c}
0\\
0\\
0\\
0
\end{array}\right),\label{eq:ExampleStationarityCondition}
\end{equation}
which is done efficiently on a classical computer once $\Omega_{t}\left[\mathbf{t'}\right]$
can be evaluated for a given set of parameters (for example, by a
Newton-Raphson method). In the case of a lattice problem, the grand
potential functional takes the following form

\begin{equation}
\Omega_{t}=\Omega'_{t'}-\frac{1}{N}\oint_{C}\frac{dz}{2\pi i}\sum_{\mathbf{\tilde{k}}}\ln\det\left[\mathbf{\hat{I}}-\mathbf{\hat{V}}\left(\mathbf{\tilde{k}}\right)\mathbf{\hat{G}'}\left(z\right)\right],\label{eq:GrandPotentialLattice}
\end{equation}
where $\mathbf{\hat{V}}\left(\mathbf{\tilde{k}}\right)$ and $\mathbf{\hat{G}'}\left(z\right)$
both depend on the chosen variational parameters (the hat notation
is explained below, it refers to the Nambu space). The contour integral
$\oint_{C}dz$ can be done as a real line integral, as a Matsubara
sum or as an efficient summation based on the continued fraction expansion
of the Fermi function \citep{Ozaki07}.

\subsubsection{The eigenvalue problem}

In order to evaluate the energy-dependent Green's function $\mathbf{\hat{G}'}\left(z\right)$,
the eigenvalue problem for one cluster

\begin{equation}
\mathcal{H}'\left|\phi_{n}\right\rangle =E_{n}\left|\phi_{n}\right\rangle \label{eq:ClusterEigenvalueProblem}
\end{equation}
must be solved for different parameters until the stationary point
is reached. For $2L_{c}$ orbitals , the eigenvalue problem of the
Hamiltonian can be solved in the occupation eigenbasis defined by

\begin{equation}
\left|n_{1\uparrow}\ldots\:n_{L_{c}\uparrow}n_{1\downarrow}\ldots\:n_{L_{c}\downarrow}\right\rangle =\prod_{i=1}^{L_{c}}\left(c_{i\uparrow}^{\dagger}\right)^{n_{i\uparrow}}\prod_{i=1}^{L_{c}}\left(c_{i\downarrow}^{\dagger}\right)^{n_{i\downarrow}}\left|\mathrm{Vac}\right\rangle ,\label{eq:OccupationBasis}
\end{equation}
where $\left|\mathrm{Vac}\right\rangle $ is the many-body vacuum.
The dimension of this Hilbert space is $4^{L_{c}}$ which means that
storing the matrices of the calculation scales prohibitively with
cluster size on a classical computer. Let's note that spatial symmetries
that commute with the cluster Hamiltonian $\mathcal{H}'$ can be used
to reduce the memory requirement of the computation \citep{Senechal08}\emph{.
} In all cases, it is useful to introduce the Nambu (singlet particle-hole)
space notation. This notation is especially useful when considering
quantum mechanical problems where an order parameter can appear from
broken gauge symmetries. In this space, field operators are replaced
by a vector $\mathbf{\Psi}_{i}^{\dagger}=\left(\begin{array}{cc}
c_{i\uparrow}^{\dagger} & c_{i\downarrow}\end{array}\right)$ such that the energy-dependent Green's function of a cluster can
be represented in the form

\begin{equation}
\begin{array}{ccl}
\mathbf{\hat{G}'}\left(\omega\right) & \equiv & \left\langle \mathbf{\Psi}\mathbf{\Psi}^{\dagger}\right\rangle _{\omega}\\
\\
 & = & \left(\begin{array}{cc}
\mathbf{G}'\left(\omega\right) & \mathbf{F}'\left(\omega\right)\\
\mathbf{F}'^{\dagger}\left(\omega\right) & -\mathbf{G}'\left(-\omega\right)
\end{array}\right),
\end{array}\label{eq:NambuGreensFunction}
\end{equation}
where the elements $G'_{ij}\left(\omega\right)=\left\langle c_{i\uparrow}c_{j\uparrow}^{\dagger}\right\rangle _{\omega}$
are the components of the single-particle Green's function and $F'_{ij}\left(\omega\right)=\left\langle c_{i\uparrow}c_{j\downarrow}\right\rangle _{\omega}$
are the components of the anomalous Green's function. The $\left\langle \ldots\right\rangle _{\omega}$notation
corresponds to the frequency-dependent correlation function (i.e.
the Fourier transformed two-point time correlation function). In
the 4-site example, these matrices would have the form \begin{widetext}

\begin{equation}
\mathbf{G}'\left(\omega\right)=\left(\begin{array}{cccc}
\left\langle c_{1\uparrow}c_{1\uparrow}^{\dagger}\right\rangle _{\omega} & \left\langle c_{1\uparrow}c_{2\uparrow}^{\dagger}\right\rangle _{\omega} & \left\langle c_{1\uparrow}c_{3\uparrow}^{\dagger}\right\rangle _{\omega} & \left\langle c_{1\uparrow}c_{4\uparrow}^{\dagger}\right\rangle _{\omega}\\
\left\langle c_{2\uparrow}c_{1\uparrow}^{\dagger}\right\rangle _{\omega} & \left\langle c_{2\uparrow}c_{2\uparrow}^{\dagger}\right\rangle _{\omega} & \left\langle c_{2\uparrow}c_{3\uparrow}^{\dagger}\right\rangle _{\omega} & \left\langle c_{2\uparrow}c_{4\uparrow}^{\dagger}\right\rangle _{\omega}\\
\left\langle c_{3\uparrow}c_{1\uparrow}^{\dagger}\right\rangle _{\omega} & \left\langle c_{3\uparrow}c_{2\uparrow}^{\dagger}\right\rangle _{\omega} & \left\langle c_{3\uparrow}c_{3\uparrow}^{\dagger}\right\rangle _{\omega} & \left\langle c_{3\uparrow}c_{4\uparrow}^{\dagger}\right\rangle _{\omega}\\
\left\langle c_{4\uparrow}c_{1\uparrow}^{\dagger}\right\rangle _{\omega} & \left\langle c_{4\uparrow}c_{2\uparrow}^{\dagger}\right\rangle _{\omega} & \left\langle c_{4\uparrow}c_{3\uparrow}^{\dagger}\right\rangle _{\omega} & \left\langle c_{4\uparrow}c_{4\uparrow}^{\dagger}\right\rangle _{\omega}
\end{array}\right)\label{eq:ExampleSingleParticleCorrelationFunction}
\end{equation}
and

\begin{equation}
\mathbf{F}'\left(\omega\right)=\left(\begin{array}{cccc}
\left\langle c_{1\uparrow}c_{1\downarrow}\right\rangle _{\omega} & \left\langle c_{1\uparrow}c_{2\downarrow}\right\rangle _{\omega} & \left\langle c_{1\uparrow}c_{3\downarrow}\right\rangle _{\omega} & \left\langle c_{1\uparrow}c_{4\downarrow}\right\rangle _{\omega}\\
\left\langle c_{2\uparrow}c_{1\downarrow}\right\rangle _{\omega} & \left\langle c_{2\uparrow}c_{2\downarrow}\right\rangle _{\omega} & \left\langle c_{2\uparrow}c_{3\downarrow}\right\rangle _{\omega} & \left\langle c_{2\uparrow}c_{4\downarrow}\right\rangle _{\omega}\\
\left\langle c_{3\uparrow}c_{1\downarrow}\right\rangle _{\omega} & \left\langle c_{3\uparrow}c_{2\downarrow}\right\rangle _{\omega} & \left\langle c_{3\uparrow}c_{3\downarrow}\right\rangle _{\omega} & \left\langle c_{3\uparrow}c_{4\downarrow}\right\rangle _{\omega}\\
\left\langle c_{4\uparrow}c_{1\downarrow}\right\rangle _{\omega} & \left\langle c_{4\uparrow}c_{2\downarrow}\right\rangle _{\omega} & \left\langle c_{4\uparrow}c_{3\downarrow}\right\rangle _{\omega} & \left\langle c_{4\uparrow}c_{4\downarrow}\right\rangle _{\omega}
\end{array}\right).\label{eq:ExampleAnomalousCorrelationFunction}
\end{equation}
\end{widetext} Methods to evaluate $\mathbf{\hat{G}'}\left(\omega\right)$
on classical and quantum computer are given in section \ref{sec:EigenvalueProblem}.

Let's notice that in the $L_{c}=2\times2$ cluster, 32 different
correlation functions have to be evaluated. In the general case, the
number of correlation functions simply scales as $4\cdot L_{c}^{2}$,
which is much smaller than the exponential scaling required for storing
the full density matrix. See subsection\ref{sub:ClassicalComputer}
and subsection \ref{sub:QuantumComputer} for the procedure to obtain
these Green's functions. 

At this point, the CPT potential in the reciprocal superlattice basis
can also be defined as

\begin{equation}
\mathbf{\hat{V}\left(\mathbf{\tilde{k}}\right)\equiv\mathbf{\hat{t}\left(\mathbf{\tilde{k}}\right)}-\hat{t}'},\label{eq:ClusterPerturbationPotential}
\end{equation}
where $\mathbf{\hat{t}\left(\mathbf{\tilde{k}}\right)}$ contains
all the one-body terms of the bare lattice $\Gamma$ (i.e. no interaction
terms) of the Hamiltonian (\ref{eq:FermiHubbardHamiltonian}). For
the example of the square lattice, this gives

\begin{equation}
\mathbf{\hat{t}\left(\mathbf{\tilde{k}}\right)}=\left(\begin{array}{cc}
\mathbf{A\left(\mathbf{\tilde{k}}\right)} & \mathbf{0}\\
\mathbf{0} & \mathbf{-A\left(\mathbf{\tilde{k}}\right)}
\end{array}\right),\label{eq:OneBodyTermLattice}
\end{equation}
where

\begin{equation}
\mathbf{A\left(\mathbf{\tilde{k}}\right)}=\left(\begin{array}{cccc}
-\mu & \epsilon\left(\tilde{k}_{x}\right) & \epsilon\left(\tilde{k}_{y}\right) & 0\\
\epsilon^{*}\left(\tilde{k}_{x}\right) & -\mu & 0 & \epsilon\left(\tilde{k}_{y}\right)\\
\epsilon^{*}\left(\tilde{k}_{y}\right) & 0 & -\mu & \epsilon\left(\tilde{k}_{x}\right)\\
0 & \epsilon^{*}\left(\tilde{k}_{y}\right) & \epsilon^{*}\left(\tilde{k}_{x}\right) & -\mu
\end{array}\right)\label{eq:AMatrix}
\end{equation}
and the dispersion relation for the square lattice is

\begin{equation}
\epsilon\left(\tilde{k}\right)=-t\left(1+e^{-2i\tilde{k}a}\right).\label{eq:DispersionRelation}
\end{equation}
The $\mathbf{\hat{t}'}$ term in equation (\ref{eq:ClusterPerturbationPotential})
contains all one-body terms of a cluster (\ref{eq:HamiltonianOneCluster}),
including the variational terms. In the example,

\begin{equation}
\mathbf{\hat{t}'}=\left(\begin{array}{cc}
\mathbf{B} & \mathbf{C}\\
\mathbf{C} & \mathbf{D}
\end{array}\right),\label{eq:OneBodyTermCluster}
\end{equation}
where

\begin{equation}
\mathbf{\mathbf{B}}=\left(\begin{array}{cccc}
-\mu'+M' & -t & -t & 0\\
-t & -\mu'-M' & 0 & -t\\
-t & 0 & -\mu'-M' & -t\\
0 & -t & -t & -\mu'+M'
\end{array}\right)\label{eq:BMatrix}
\end{equation}
and
\begin{equation}
\mathbf{\mathbf{D}}=\left(\begin{array}{cccc}
\mu'+M' & t & t & 0\\
t & \mu'-M' & 0 & t\\
t & 0 & \mu'-M' & t\\
0 & t & t & \mu'+M'
\end{array}\right).\label{eq:DMatrix}
\end{equation}
The pairing part is given by 
\begin{equation}
\mathbf{\mathbf{C}}=\left(\begin{array}{cccc}
\Delta' & \Delta_{d}' & -\Delta_{d}' & 0\\
\Delta_{d}' & \Delta' & 0 & -\Delta_{d}'\\
-\Delta_{d}' & 0 & \Delta' & \Delta_{d}'\\
0 & -\Delta_{d}' & \Delta_{d}' & \Delta'
\end{array}\right).\label{eq:CMatrix}
\end{equation}

\subsubsection{The lattice-perturbed Green's function}

Once the saddle point $\mathbf{t}_{*}=\left(\begin{array}{c}
\mu'_{*}\\
\Delta'_{*}\\
\Delta'_{d*}\\
M'_{*}
\end{array}\right)$ of equation (\ref{eq:GeneralStationarityCondition}) is found, the
function $\mathbf{\hat{G}'}\left(\omega,\mathbf{t}_{*}\right)$ and
$\hat{\mathbf{V}}\left(\mathbf{\tilde{k}},\mathbf{t}_{*}\right)$
are evaluated and the lattice-perturbed Green's function can be computed.
From here the dimensionality of the matrices involved in the calculations
scales only as the square of the number of orbitals and can be performed
easily on a classical computer. The lattice-perturbed Green's function
can be calculated to first order as

\begin{equation}
\begin{array}{rcl}
\mathbf{\mathcal{\hat{G}}\left(\mathbf{\mathbf{\tilde{k}},\omega}\right)} & = & \left(\mathbf{\hat{G}'^{-1}}\left(\omega\right)-\hat{\mathbf{V}}\left(\mathbf{\tilde{k}}\right)\right)^{-1}\\
\\
 & = & \left(\begin{array}{cc}
\mathbf{\mathcal{G}}'\left(\mathbf{\mathbf{\tilde{k}}},\omega\right) & \mathbf{\mathcal{F}}'\left(\mathbf{\mathbf{\tilde{k}}},\omega\right)\\
\mathbf{\mathcal{F}}'^{\dagger}\left(\mathbf{\mathbf{\tilde{k}}},\omega\right) & -\mathbf{\mathcal{G}}'\left(\mathbf{\mathbf{\tilde{k}}},-\omega\right)
\end{array}\right).
\end{array}\label{eq:LatticePerturbedGreensFunction}
\end{equation}
Note that the $\mathbf{\mathcal{G}}$ and $\mathbf{\mathcal{F}}$
matrices have dimension $L_{c}\times L_{c}$. At this point the problem
is solved and many observable quantities can be computed efficiently
\citep{Rickayzen91}.

\subsection{Calculation of observables\label{sub:CalculationOfObservables}}

Based on \citep{Kaneko14}, this subsection contains examples of observables
useful in explaining the result of experiments and landmark properties
of the FHM. 

The average particle density is

\begin{equation}
n=\left\langle n_{i\sigma}\right\rangle =\frac{1}{NL_{c}}\oint_{C}\frac{dz}{2\pi i}\sum_{\mathbf{\tilde{k}}}\sum_{i=1}^{L_{c}}\mathbf{\mathcal{G}}_{ii}\left(\mathbf{\mathbf{\tilde{k}}},\omega\right)\label{eq:AverageParticleDensity}
\end{equation}
and must agree with the value given by (\ref{eq:ThermodynamicalConsistencyOccupationNumber}).
The chemical potential $\mu$ can be scanned until a desired value
of $n$ is found. For superconducting problem in the FHM, it is useful
to fix the chemical potential $\mu$ such that the lattice is maintained
at quarter filling $n=0.25$. The superconducting gap is given by

\begin{equation}
\Delta=\left\langle c_{i\uparrow}c_{j\downarrow}\right\rangle =\frac{1}{NL_{c}}\oint_{C}\frac{dz}{2\pi i}\sum_{\mathbf{\tilde{k}}}\sum_{i=1}^{L_{c}}\mathbf{\mathcal{F}}_{ii}\left(\mathbf{\mathbf{\tilde{k}}},\omega\right).\label{eq:AnomalousExpectationValue}
\end{equation}

\emph{} To recover the Green's functions of the full lattice $\Gamma$,
the ``clustering'' (which is a unitary transformation) is undone
and, taking into account the artificial translational symmetry breaking
of the lattice, the single-particle and anomalous CPT Green's functions
are recovered in the lattice reciprocal space

\begin{equation}
\begin{array}{ccl}
\mathcal{G}_{\mathrm{cpt}}\left(\mathbf{k,\omega}\right) & = & \frac{1}{L_{c}}\sum_{i,j=1}^{L_{c}}\mathcal{G}_{ij}\left(\mathbf{k},\omega\right)e^{-i\mathbf{k}\cdot\left(\mathbf{r}_{i}-\mathbf{r}_{j}\right)}\\
\\
\mathcal{F}_{\mathrm{cpt}}\left(\mathbf{k,\omega}\right) & = & \frac{1}{L_{c}}\sum_{i,j=1}^{L_{c}}\mathcal{F}_{ij}\left(\mathbf{k},\omega\right)e^{-i\mathbf{k}\cdot\left(\mathbf{r}_{i}-\mathbf{r}_{j}\right)}.
\end{array}\label{eq:CPTGreensFunction}
\end{equation}
From these quantities, the single-particle quasiparticle spectrum
and the Bogoliubov quasiparticle spectrum can be evaluated as

\begin{equation}
\begin{array}{ccl}
A\left(\mathbf{k,\omega}\right) & = & -\frac{1}{\pi}\lim_{\eta\rightarrow0^{+}}\mathrm{Im}\,\mathcal{G}_{\mathrm{cpt}}\left(\mathbf{k},\omega+i\eta\right)\\
\\
F\left(\mathbf{k,\omega}\right) & = & -\frac{1}{\pi}\lim_{\eta\rightarrow0^{+}}\mathrm{Im}\,\mathcal{F}_{\mathrm{cpt}}\left(\mathbf{k},\omega+i\eta\right),
\end{array}\label{eq:QuasiparticleSpectrum}
\end{equation}
from which the density of states is found to be 

\begin{equation}
N\left(\omega\right)=\frac{1}{N}\sum_{\mathbf{k}}A\left(\mathbf{k},\omega\right).\label{eq:DensityOfStates}
\end{equation}

The Fermion momentum distribution and the condensation amplitude momentum
distribution are respectively given by

\begin{equation}
\begin{array}{ccl}
N\left(\mathbf{k}\right) & = & \oint_{C}\frac{dz}{2\pi i}\mathcal{G}_{\mathrm{cpt}}\left(\mathbf{k},z\right)\\
\\
F\left(\mathbf{k}\right) & = & \oint_{C}\frac{dz}{2\pi i}\mathcal{F}_{\mathrm{cpt}}\left(\mathbf{k},z\right).
\end{array}\label{eq:MomentumDistribution}
\end{equation}
For the case of a lattice with superconductivity, an interesting observable
is the pair coherence length in real and reciprocal space given by

\begin{equation}
\xi^{2}=\frac{\sum_{\mathbf{r}}\mathbf{r}^{2}\left|F\left(\mathbf{r}\right)\right|^{2}}{\sum_{\mathbf{r}}\left|F\left(\mathbf{r}\right)\right|^{2}}=\frac{\sum_{\mathbf{k}}\left|\mathbf{\nabla_{k}}F\left(\mathbf{k}\right)\right|^{2}}{\sum_{\mathbf{k}}\left|F\left(\mathbf{k}\right)\right|^{2}}.\label{eq:PairCoherenceLength}
\end{equation}

Depending on the problem, more observable can be computed with similar
methods. Note also that the contour integrals map to the following
form in the real time domain 
\begin{equation}
\oint_{C}\frac{dz}{2\pi i}\mathcal{G}^{R}\left(z\right)\rightarrow\int_{-\infty}^{\infty}d\omega f\left(\omega\right)\mathcal{G}^{R}\left(\omega\right)
\end{equation}
in the case where the retarded part of the Green's function is used
to compute the integral. The Fermi function has the usual form $f\left(\omega\right)=\frac{1}{1+e^{\frac{\mu-\omega}{T}}}$.
The self-energy variational approach has been outlined and the method
which starts with a Hubbard-like description of the microscopic details
of a given solid and compute its thermodynamic properties in a systematic
way is complete. The next section reviews how the eigenvalue problem
(\ref{eq:ClusterEigenvalueProblem}) is typically solved on classical
computers and introduces the quantum subroutine.

\section{Solving the eigenvalue problem on a quantum computer\label{sec:EigenvalueProblem}}

Solving the eigenvalue problem (\ref{eq:ClusterEigenvalueProblem})
for a large number of electrons is exponentially costly in memory
as the number of orbitals increases. This section is divided the following
way. First the classical eigenvalue solver for the Green's function
is described. Then the Jordan-Wigner transformation is used to map
the cluster Hamiltonian to a quantum register. A method to generate
initial Gibbs states in a quantum computer in reviewed and finally
a procedure to extract the Green's function out of the Gibbs state
is explained. The full quantum procedure is shown to be efficient
in quantum memory resources.

\subsection{The method on a classical computer\label{sub:ClassicalComputer}}

The resource intensive part of the numerical variation solver is the
computation of the energy-dependent Green's function of the cluster
$\mathbf{\hat{G}'}\left(\omega,\mathbf{t}\right)$. On a classical
computer, the memory used to store the description of the state of
the system scales exponentially in system size.

\begin{table}
\begin{centering}
\begin{tabular}{|c|c|}
\hline 
Number of orbitals & Memory required\tabularnewline
\hline 
\hline 
3 & 1 KB\tabularnewline
\hline 
8 & 1 MB\tabularnewline
\hline 
13 & 1 GB\tabularnewline
\hline 
18 & 1 TB\tabularnewline
\hline 
23 & 1 PB\tabularnewline
\hline 
\end{tabular}
\par\end{centering}

\caption{Order of magnitude estimation of the classical memory required to
store the full finite temperature density matrix of a cluster with
a given number of irreducible orbitals for a general cluster Hamiltonian.
It is assumed that each matrix element is stored as a complex double-precision
number (16 bytes/element) and no optimization is used.\label{tab:ClassicalMemory}}

\end{table}

Typically, the Hamiltonian (\ref{eq:HamiltonianOneCluster}) is encoded
in the occupation basis (\ref{eq:OccupationBasis}) and the Schrödinger
equation (\ref{eq:ClusterEigenvalueProblem}) is solved explicitly
using an appropriate numerical diagonalization method. As shown in
table \ref{tab:ClassicalMemory}, the memory usage scales exponentially
with system size and diagonalization typically scales as $O\left(L_{c}^{3}\right)$
in the number of arithmetic operation required. When successful, a
set of eigenvalues $\left\{ E_{n}\right\} $ and associated eigenstates
$\left\{ \left|\phi_{n}\right\rangle \right\} $ are obtained. If
the cluster has $L_{c}$ sites with 2 electrons each (spin up and
down), then there are $4^{L_{c}}$ eigenstates. The rest of the procedure
is the following:
\begin{enumerate}
\item Write $\omega_{mn}=E_{n}-E_{m}$.
\item Write the occupation probabilities $P_{mn}=\frac{e^{-\beta E_{n}}+e^{-\beta E_{m}}}{Z}$.
Note that $\beta\equiv T^{-1}$ is the inverse temperature and $Z=\mathrm{Tr\,e^{-\beta\mathcal{H}'}}$
is the partition function.
\item Define the electron-like and hole-like amplitude $Q_{imn}^{^{\left(e\uparrow\right)}}=\left\langle \phi_{m}\right|c_{i\uparrow}\left|\phi_{n}\right\rangle $
and $Q_{imn}^{^{\left(h\downarrow\right)}}=\left\langle \phi_{m}\right|c_{i\downarrow}^{\dagger}\left|\phi_{n}\right\rangle $
.
\item Vectorize the $m,n\longrightarrow r$ indices to obtain the matrices
$\hat{E}_{rs}=\delta_{rs}\omega_{r}$ and $\hat{\Pi}_{rs}=\delta_{rs}P_{r}$
. The amplitude matrices then take the form
\begin{equation}
\mathbf{\hat{Q}}=\left(\begin{array}{c}
Q_{1r}^{^{\left(e\uparrow\right)}}\\
\vdots\\
Q_{L_{c}r}^{^{\left(e\uparrow\right)}}\\
Q_{1r}^{^{\left(h\downarrow\right)}}\\
\vdots\\
Q_{L_{c}r}^{^{\left(h\downarrow\right)}}
\end{array}\right)\label{eq:QAmplitudeMatrix}
\end{equation}
and can be recast as $\mathbf{\hat{Q}}'=\mathbf{\hat{Q}}\sqrt{\hat{\Pi}}$
at non-zero temperature. It is also useful to define and compute $\hat{\mathbf{g}}\left(\omega\right)=\frac{\hat{\mathbf{1}}}{\omega-\mathbf{\hat{E}}}$.
It an be noted that $\hat{\mathbf{Q}}$ is a $2L_{c}\times16^{L_{c}}$
matrix which scales exponentially in memory with the size of the system
being studied.
\item Then compute (\ref{eq:NambuGreensFunction}) as $\mathbf{\hat{G}'}\left(\omega\right)=\mathbf{\hat{Q}}'\hat{\mathbf{g}}\left(\omega\right)\mathbf{\hat{Q}}'^{\dagger}$.
This is the most time-consuming step on a classical computer, especially
at non-zero temperature.
\item The grand potential functional (\ref{eq:GrandPotentialLattice}) and
the lattice-perturbed Green's function (\ref{eq:LatticePerturbedGreensFunction})
can then be evaluated to respectively solve the saddle-point problem
and compute observables.
\end{enumerate}

\subsection{The method on a quantum computer\label{sub:QuantumComputer}}

Computing the Green's function of the cluster $\mathbf{\hat{G}'}\left(\omega,\mathbf{t}\right)$
on a quantum computer is possible in a hybrid analog-digital simulator.
The first step generates a Gibbs state $\rho_{\mathrm{Gibbs}}\left(T\right)$
with some temperature $T$ (or $\beta=\frac{1}{T}$) measured on the
digital register and the second step measures the correlation function
of the cluster on an analog channel. The Jordan-Wigner transformation
is used to map the Fermi-Hubbard Hamiltonian to a quantum register.
The general procedure is the following:
\begin{enumerate}
\item \emph{Map the cluster Hamiltonian (\ref{eq:HamiltonianOneCluster})
to a qubit system with the Jordan-Wigner transformation.}
\item \emph{Evaluate the two-point correlation functions (\ref{eq:NambuGreensFunction})
for many different times for at least a full Hamiltonian cycle (at
zero temperature) or until correlations flatten out. Fourier transform
to obtain the frequency-dependent correlation functions. The Hamiltonian
is evolved in time using Trotter steps. Note that in the Jordan-Wigner
basis, $O\left(L_{c}\right)$ gates are needed at each time step.
The full density matrix does not need to be measured, only $O\left(L_{c}^{2}\right)$
correlation functions need to be evaluated. }
\item \emph{Again, the grand potential functional (\ref{eq:GrandPotentialLattice})
and the lattice-perturbed Green's function (\ref{eq:LatticePerturbedGreensFunction})
can then be evaluated efficiently on a classical computer (simple
linear algebra on small $2L_{c}\times2L_{c}$ matrices) to respectively
solve the saddle-point problem and compute observable.}
\end{enumerate}
\begin{figure}
\begin{centering}
\includegraphics[width=3.375in]{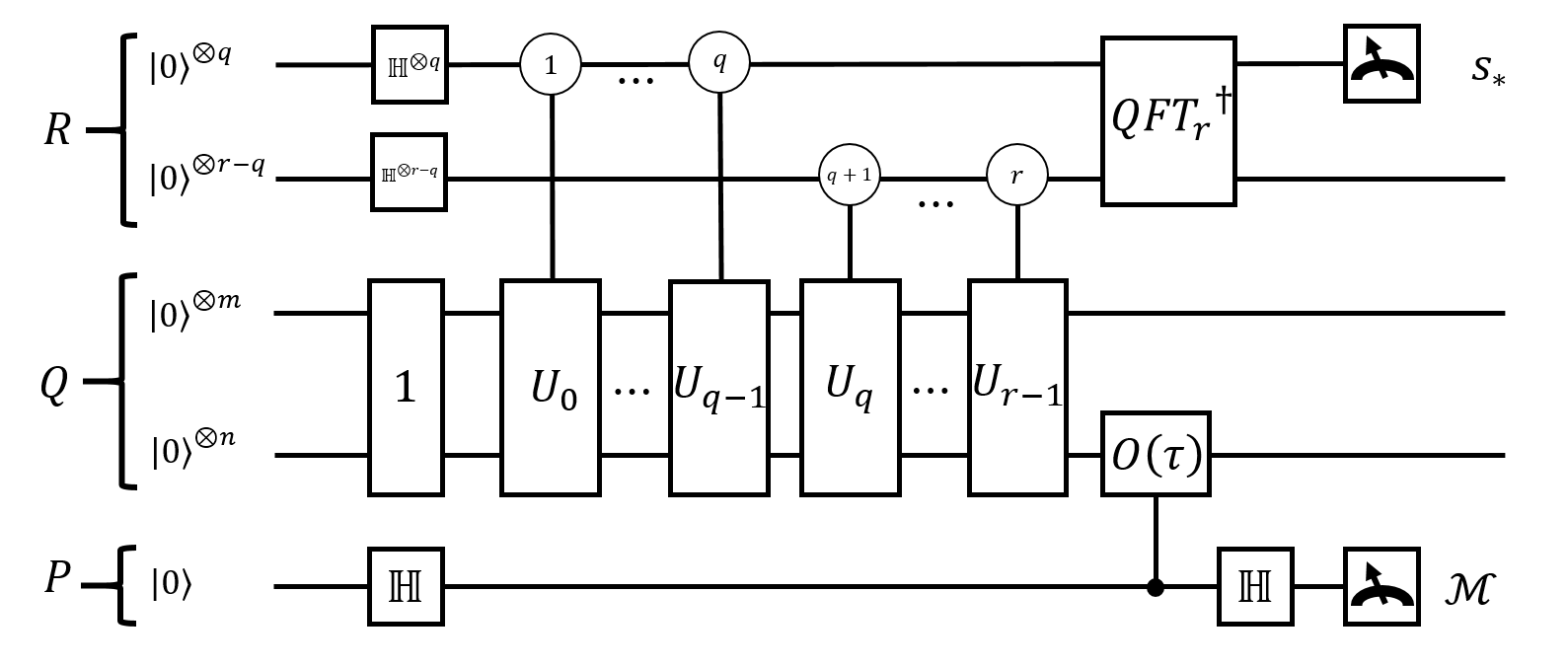}
\par\end{centering}

\caption{Circuit to simulate the time-dependent correlation function (\ref{eq:MeasuredCorrelationFunction})
of the cluster Hamiltonian (\ref{eq:HamiltonianOneCluster}). The
first part meant to generate a Gibbs state is taken from \citep{Riera12}.
Register $R$ is used in the modified phase estimation scheme to prepare
a rectangular state between the bath and the system contained in register
$Q$. When the bath is traced out the system channel is left in a
Gibbs state from which the different correlation functions can be
read from the one-qubit register $P$. The size of register $Q$ depends
on the number of orbitals in the simulated cluster (typically $n=2L_{c}$)
and the bath size (which can be some constant factor larger than the
system register). Register $R$ is used as a digital component and
$q$ is therefore be the size required for the desired floating point
accuracy on reading $s_{*}$. Note that the numbers in the controlled
gates of register $R$ denote the index of the qubit which is acting
as the control.\label{fig:FullCircuit}}
\end{figure}

A full quantum circuit to measure $\mathbf{\hat{G}'}\left(\omega,\mathbf{t}\right)$
is shown in figure \ref{fig:FullCircuit}. The specific algorithm
\citep{Riera12} to create a Gibbs state was chosen mostly for aesthetic
reasons. It appears to be the only Gibbs state generation method that
provides bounds on all parameters of the algorithm and that can be
written in a circuit model. For completeness the main results of \citep{Riera12}
are summarized and commented in appendix \ref{sec:GibbsStatePreparation}.
There is no reason to believe that other sampling methods \citep{Poulin09,Bilgin10,Temme11}
would not work also. A variational eigensolver \citep{Peruzzo14}
or an adiabatic quantum algorithm \citep{Wecker15} could hypothetically
be used to supply the initial ground state in the case of a simulation
at zero temperature. 

Equation (\ref{eq:ClusterEigenvalueProblem}) does not need to be
solved explicitly on a quantum computer, only a few correlation functions
of interest need to be computed, this is explained in details in subsection
\ref{sub:MeasuringCorrelationFunction}. The controlled evolution
gates shown in figure \ref{fig:FullCircuit} assume that the Hamiltonian
of the cluster can be mapped to a Hamiltonian in the quantum computer
Hilbert space. Here is the procedure to make the mapping that requires
no oracle black box for $\mathcal{H}'$. The Hamiltonian (\ref{eq:HamiltonianOneCluster})
is broken into $M$ non-commuting parts such that

\begin{equation}
\mathcal{H}'=\sum_{i=1}^{M}\mathcal{H}_{i}'.\label{eq:TrotterHamiltonian}
\end{equation}
Each time-step $\Delta\tau$ evolution of the cluster Hamiltonian
\citep{LasHeras15} can be simulated with $n_{T}$ Trotter-Suzuki
steps

\begin{equation}
e^{-i\mathcal{H}'\Delta\tau}\simeq\left(\prod_{i=1}^{M}e^{-\frac{i\mathcal{H}_{i}'\Delta\tau}{n_{T}}}\right)^{n_{T}}+\sum_{i<j}\frac{\left[\mathcal{H}_{i}',\mathcal{H}_{j}'\right]\Delta\tau^{2}}{2n_{T}}+\ldots.\label{eq:TrotterFormula}
\end{equation}
The size of those time-steps set the upper bound in the simulated
energy spectrum which scales as $\omega_{\mathrm{max}}\propto\frac{1}{\Delta\tau}$,
while the lowest energy scales at the inverse of the total simulation
time. 

The creation and annihilation operators of the Hamiltonian can be
mapped to the quantum computational basis using a Jordan-Wigner transformation
\citep{Nielsen05}. If there are $2L_{c}$ electrons, then the Jordan-Wigner
\citep{Nielsen05} transformed creation operators are given by

\begin{equation}
\begin{array}{ccl}
c_{i\uparrow}^{\dagger} & = & \mathbb{I}^{\otimes2L_{c}-i}\otimes\sigma_{+}\otimes\sigma_{z}^{\otimes i-1}\\
\\
c_{i\downarrow}^{\dagger} & = & \mathbb{I}^{\otimes L_{c}-i}\otimes\sigma_{+}\otimes\sigma_{z}^{\otimes L_{c}+i-1}
\end{array}.\label{eq:JordanWignerMap}
\end{equation}
In this notation,

\begin{equation}
\mathbb{\sigma}^{\otimes k}\equiv\begin{cases}
1 & k=0\\
\sigma & k=1\\
\sigma\otimes\mathbb{\sigma}^{\otimes k-1} & k>1
\end{cases},\label{eq:TensorExponentNotation}
\end{equation}
also $\sigma_{+}=\frac{\left(\sigma_{x}+i\sigma_{y}\right)}{2}$,
$\sigma_{-}=\sigma_{+}^{\dagger}$ and $\sigma_{z}=2\sigma_{n}-\mathbb{I}$,
where $\sigma_{n}\equiv\sigma_{+}\sigma_{-}$. The relations $\sigma_{+}\sigma_{z}=\sigma_{+}=-\sigma_{z}\sigma_{+}$
and $\sigma_{z}\sigma_{-}=\sigma_{-}=-\sigma_{-}\sigma_{z}$ can also
be used. Note that the Jordan-Wigner transformation is independent
of the Hamiltonian of the system and the dimensionality of the system.
\begin{widetext}In the Pauli basis of the quantum computer, the terms
of the cluster Hamiltonian (\ref{eq:HamiltonianOneCluster}) transform
to

\begin{equation}
\begin{array}{rcl}
-t\sum_{\sigma}\left(c_{i\sigma}^{\dagger}c_{j\sigma}+c_{j\sigma}^{\dagger}c_{i\sigma}\right) & \longrightarrow & -t\left(\mathbb{I}^{\otimes L_{c}}\otimes\mathbb{T}_{L_{c}}\left(i,j\right)+\mathbb{T}_{L_{c}}\left(i,j\right)\otimes\mathbb{I}^{\otimes L_{c}}\right)\\
\\
-\mu'\sum_{\sigma}n_{i\sigma} & \longrightarrow & -\mu'\left(\mathbb{I}^{\otimes L_{c}}\otimes\mathbb{T}_{L_{c}}\left(i\right)+\mathbb{T}_{L_{c}}\left(i\right)\otimes\mathbb{I}^{\otimes L_{c}}\right)\\
\\
Un_{i\uparrow}n_{i\downarrow} & \longrightarrow & U\left(\mathbb{T}_{L_{c}}\left(i\right)\otimes\mathbb{T}_{L_{c}}\left(i\right)\right)\\
\\
\Delta'\left(c_{i\uparrow}^{\dagger}c_{i\downarrow}^{\dagger}+c_{i\downarrow}c_{i\uparrow}\right) & \longrightarrow & \Delta'\mathbb{D}_{L_{c}}\left(i,i\right).
\end{array}\label{eq:ExamplesJordanWigner}
\end{equation}
The strings of Pauli matrices are defined as

\begin{equation}
\mathbb{T}_{L_{c}}\left(i,j\right)\equiv\mathbb{I}^{\otimes L_{c}-i}\otimes\left(\sigma_{+}\otimes\sigma_{z}^{\otimes i-j-1}\otimes\sigma_{-}+\sigma_{-}\otimes\sigma_{z}^{\otimes i-j-1}\otimes\sigma_{+}\right)\otimes\mathbb{I}^{\otimes j-1}\label{eq:KinOperator}
\end{equation}
where $i>j$ between 1 and $L_{c}$ and

\begin{equation}
\mathbb{T}_{L_{c}}\left(i\right)\equiv\mathbb{I}^{\otimes L_{c}-i}\otimes\sigma_{n}\otimes\mathbb{I}^{\otimes i-1}.\label{eq:LocOperator}
\end{equation}
Since $\mathbb{T}_{L_{c}}\left(i,j\right)$ and $\mathbb{T}_{L_{c}}\left(i\right)$
conserve total spin in the Pauli basis, they are also number conserving
in the occupation basis. For pairing terms it is also useful to define

\begin{equation}
\mathbb{D}_{L_{c}}\left(i,j\right)\equiv\mathbb{I}^{\otimes L_{c}-j}\otimes\left(\sigma_{+}\otimes\sigma_{z}^{\otimes L_{c}-i+j-1}\otimes\sigma_{+}+\sigma_{-}\otimes\sigma_{z}^{\otimes L_{c}-i+j-1}\otimes\sigma_{-}\right)\otimes\mathbb{I}^{\otimes i-1}.\label{eq:IntOperator}
\end{equation}
In this case, $i$ and $j$ can be anything between 1 and $L_{c}$.
The terms of $\mathbb{D}_{L_{c}}\left(i,j\right)$ do not conserve
total spin in the Pauli basis as they do not conserve the total number
of particles in the occupation basis.

\end{widetext}

In cases where the number of electrons in conserved in the cluster
Hamiltonian (with superconductivity, the anomalous pairing terms break
this symmetry), it is possible use a Bravyi-Kitaev transformation
\citep{Seeley12} for an improvement in the quantum memory usage of
the algorithm ($O\left(\ln\,L_{c}\right)$). The mapping of $\mathcal{H}'$
to the quantum computer is known and a method to generate Gibbs state
has been chosen, the correlation functions can be measured.

\subsection{Measuring the correlation function\label{sub:MeasuringCorrelationFunction}}

\begin{figure}
\begin{centering}
\includegraphics[width=3.375in]{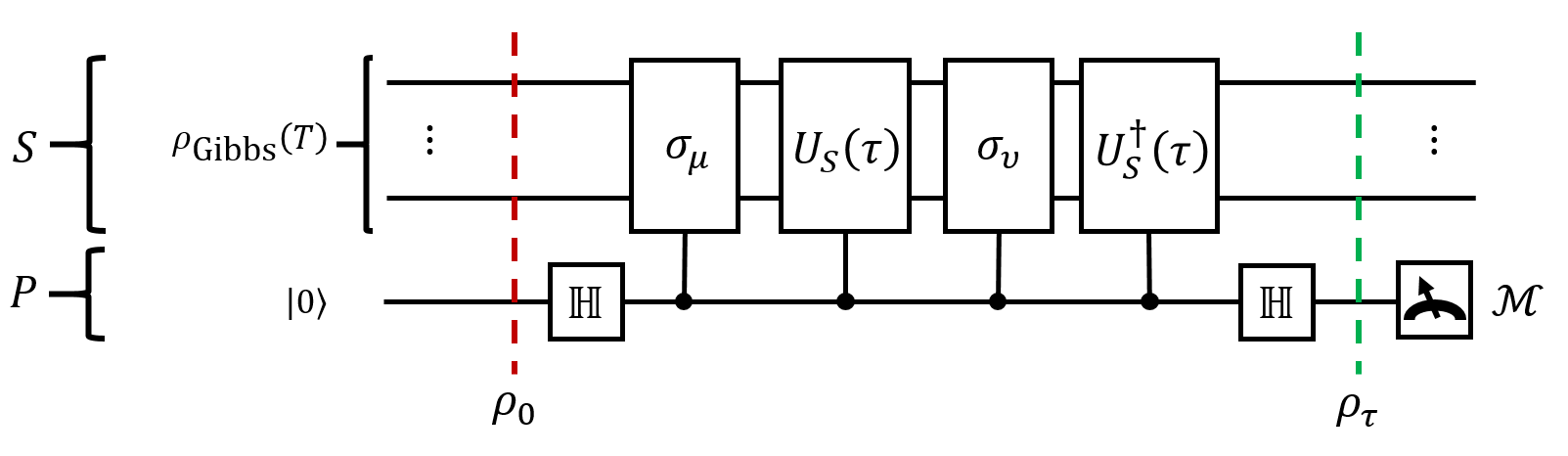}
\par\end{centering}

\caption{Circuit to measure the correlation function (\ref{eq:MeasuredCorrelationFunction})
from an input Gibbs state. Register $S$ initially contains a given
Gibbs state at inverse temperature $\beta$ and register $P$ is a
single qubit initialized in the zero state. $P$ is put in a state
superposition by applying a Hadamard gate $\mathbb{H}$ and then used
to apply the controlled evolution sequence $O_{\mu\nu}\left(\tau\right)\equiv U_{S}^{\dagger}\left(\tau\right)\sigma_{\nu}U_{S}\left(\tau\right)\sigma_{\mu}$
with $U_{S}\left(\tau\right)=e^{-i\mathcal{H}'\tau}$ to the system
channel. Finally the state superposition is reversed by a last Hadamard
gate and the measurement in repeated to obtain the probability $P\left(\mathcal{M}\right)$,
which returns information on the cluster Green's function (\ref{eq:NambuGreensFunction}).
\label{fig:CorrelationCircuit}}
\end{figure}

In this section an analog circuit is used to measure the correlation
functions of a cluster Hamiltonian at some temperature $T$ using
a variation of the phase estimation algorithm is explained \citep{Abrams97}..
The Nambu single-particle Green's function of the cluster $\mathbf{\hat{G}'}\left(\omega,\mathbf{t}\right)$
can then be recovered from the correlation function. The quantum circuit
is shown in figure \ref{fig:CorrelationCircuit}. It is a variation
on DQC1 (deterministic quantum computation with one quantum bit) \citep{Knill98,Datta08}
and phase estimation.

A thermal density matrix of the simulated system must first be prepared
in register $S$

\begin{equation}
\rho_{0}=\rho_{\mathrm{Gibbs}}\left(\beta\right)\otimes\left|0\right\rangle \left\langle 0\right|,\label{eq:ThermalDensityMatrix}
\end{equation}
where 
\begin{equation}
\rho_{\mathrm{Gibbs}}\left(\beta\right)\equiv\frac{1}{Z}\sum_{m}e^{-\beta E_{m}}\left|\phi_{m}\right\rangle \left\langle \phi_{m}\right|\label{eq:GibbsState}
\end{equation}
is a Gibbs state at some given temperature. It is to be expected that
preparing a low temperature Gibbs state (large $\beta$) is hard in
general \citep{Childs14}, while high temperature Gibbs states $\beta\rightarrow0$
are simply fully mixed states which are easier to prepare. 

A sequence of controlled gates and controlled Hamiltonian evolution
follows the application of a Hadamard gate on register $P$. The unitary
evolution generated by the cluster Hamiltonian (\ref{eq:HamiltonianOneCluster})
is defined as

\begin{equation}
\begin{array}{ccl}
U_{S}\left(\tau\right) & \equiv & e^{-i\mathcal{H}'\tau}\\
\\
 & = & \sum_{m}e^{-iE_{m}\tau}\left|\phi_{m}\right\rangle \left\langle \phi_{m}\right|.
\end{array}\label{eq:UnitaryOperator}
\end{equation}
For convenience of notation (as seen in figure \ref{fig:FullCircuit}),
it is useful to introduce the set of gates $O_{\mu\nu}$

\begin{equation}
O_{\mu\nu}\left(\tau\right)\equiv U_{S}^{\dagger}\left(\tau\right)\sigma_{\nu}U_{S}\left(\tau\right)\sigma_{\mu}\label{eq:SimulatedCorrelationOperator}
\end{equation}
that define the application of a self-adjoint operator $\sigma_{\nu}$
on the system (detailed below), followed by forward time evolution,
then the application of another $\sigma_{\mu}$ and finally a reverse
time evolution. When applied to a Gibbs state in a phase-estimation
circuit, the state of the computer at time $\tau$ is described by
\begin{widetext}

\begin{equation}
\begin{array}{cccccl}
\rho_{\tau} & = &  & \frac{1}{4}\left(\rho_{\mathrm{Gibbs}}+\rho_{\mathrm{Gibbs}}O_{\mu\nu}^{\dagger}\left(\tau\right)+O_{\mu\nu}\left(\tau\right)\rho_{\mathrm{Gibbs}}+O_{\mu\nu}\left(\tau\right)\rho_{\mathrm{Gibbs}}O_{\mu\nu}^{\dagger}\left(\tau\right)\right) & \otimes & \left|0\right\rangle \left\langle 0\right|\\
 &  & + & \frac{1}{4}\left(\rho_{\mathrm{Gibbs}}-\rho_{\mathrm{Gibbs}}O_{\mu\nu}^{\dagger}\left(\tau\right)+O_{\mu\nu}\left(\tau\right)\rho_{\mathrm{Gibbs}}-O_{\mu\nu}\left(\tau\right)\rho_{\mathrm{Gibbs}}O_{\mu\nu}^{\dagger}\left(\tau\right)\right) & \otimes & \left|0\right\rangle \left\langle 1\right|\\
 &  & + & \frac{1}{4}\left(\rho_{\mathrm{Gibbs}}+\rho_{\mathrm{Gibbs}}O_{\mu\nu}^{\dagger}\left(\tau\right)-O_{\mu\nu}\left(\tau\right)\rho_{\mathrm{Gibbs}}-O_{\mu\nu}\left(\tau\right)\rho_{\mathrm{Gibbs}}O_{\mu\nu}^{\dagger}\left(\tau\right)\right) & \otimes & \left|1\right\rangle \left\langle 0\right|\\
 &  & + & \frac{1}{4}\left(\rho_{\mathrm{Gibbs}}-\rho_{\mathrm{Gibbs}}O_{\mu\nu}^{\dagger}\left(\tau\right)-O_{\mu\nu}\left(\tau\right)\rho_{\mathrm{Gibbs}}+O_{\mu\nu}\left(\tau\right)\rho_{\mathrm{Gibbs}}O_{\mu\nu}^{\dagger}\left(\tau\right)\right) & \otimes & \left|1\right\rangle \left\langle 1\right|.
\end{array}\label{eq:EvolvedThermalDensityMatrix}
\end{equation}
\end{widetext}

It can be seen that $\rho_{\tau}$ contains the information of the
correlation function $\left\langle \sigma_{\mu}\left(\tau\right)\sigma_{\nu}\left(0\right)\right\rangle $,
which can be measured by evaluating the probability $P_{\mu\nu}\left(\mathcal{M}=0\left(1\right),\tau\right)$
of measuring zero (one) in register $P$ (and then Fourier transformed
to obtain $\left\langle \sigma_{\mu}\sigma_{\nu}\right\rangle _{\omega}$).
Formally, the interesting correlation functions that need to be extracted
have the textbook form \citep{Rickayzen91}

\begin{equation}
\begin{array}{ccl}
C_{\mu\nu}\left(\tau\right) & \equiv & \left\langle \sigma_{\mu}\left(\tau\right)\sigma_{\nu}\left(0\right)\right\rangle \\
\\
 & = & \mathrm{Tr}\left[\rho_{\mathrm{Gibbs}}O_{\mu\nu}^{\dagger}\left(\tau\right)+O_{\mu\nu}\left(\tau\right)\rho_{\mathrm{Gibbs}}\right]\\
\\
 & = & \sum_{m}\sum_{n}e^{-i\tau\left(E_{m}-E_{n}\right)}A_{\mu\nu}^{mn},
\end{array}\label{eq:CorrelationFunction}
\end{equation}
where $A_{\mu\nu}^{mn}\equiv\frac{e^{-\beta E_{m}}+e^{-\beta E_{n}}}{Z}\left\langle \phi_{n}\right|\sigma_{\mu}\left|\phi_{m}\right\rangle \left\langle \phi_{m}\right|\sigma_{\nu}\left|\phi_{n}\right\rangle $.
Note that these functions always outputs a real number. If the controlled
operation $c-O_{\mu\nu}\left(\tau\right)$ is applied for a time $\tau>0$,
the phase estimation algorithm yields the following probability for
the two different outcomes $\mathcal{M}=0$ and $\mathcal{M}=1$

\begin{equation}
\begin{array}{ccl}
P_{\mu\nu}\left(\mathcal{M}=0,\tau\right) & = & \frac{1}{2}\left(1+\frac{1}{2}C_{\mu\nu}\left(\tau\right)\right)\\
\\
P_{\mu\nu}\left(\mathcal{M}=1,\tau\right) & = & \frac{1}{2}\left(1-\frac{1}{2}C_{\mu\nu}\left(\tau\right)\right).
\end{array}\label{eq:PhaseEstimationProbabilities}
\end{equation}
Then from measuring the probability trajectory, the functions (\ref{eq:CorrelationFunction})
can be recovered as

\begin{equation}
C_{\mu\nu}\left(\tau\right)=2\left(P_{\mu\nu}\left(\mathcal{M}=0,\tau\right)-P_{\mu\nu}\left(\mathcal{M}=1,\tau\right)\right).\label{eq:MeasuredCorrelationFunction}
\end{equation}

As in DQC1 \citep{Knill98}, in general it is not useful to use multiple
ancillary qubits and an inverse Fourier transform to extract multiple
bits of the probabilities $P_{\mu\nu}$ at each measurement shot since
the input $\rho_{\mathrm{Gibbs}}$ is a state mixture. In the case
where the simulated temperature is so low that the input Gibbs state
is effectively is a pure (non-degenerate) ground state, it is plausible
that adding qubits to register $P$ would speed-up the measurement
of the $P_{\mu\nu}$'s in the traditional sense of phase estimation
\citep{Nielsen01}. The retarded Green's function can be computed
numerically as

\begin{equation}
G_{\mu\nu}^{R}\left(\tau\right)\equiv-i\theta\left(\tau\right)C_{\mu\nu}\left(\tau\right)\label{eq:RetardedRealTimeGreensFunction}
\end{equation}
where $\theta\left(\tau\right)$ is the Heaviside function. It can
be Fourier transformed to get the Green's function in the frequency
domain

\begin{equation}
G_{\mu\nu}^{R}\left(\omega\right)=\int_{-\infty}^{\infty}d\tau\:e^{-i\omega\tau}G_{\mu\nu}^{R}\left(\tau\right).\label{eq:RetardedFrequencyGreensFunction}
\end{equation}
The spectral function can be obtained from the retarded Green's function
as

\begin{equation}
\begin{array}{ccl}
A_{\mu\nu}\left(\omega\right) & = & \frac{i}{2\pi}\left(G_{\mu\nu}^{R}\left(\omega\right)-G_{\mu\nu}^{A}\left(\omega\right)\right)\\
\\
 & = & -\frac{1}{\pi}\mathrm{Im}\left\{ G_{\mu\nu}^{R}\left(\omega\right)\right\} .
\end{array}\label{eq:SpectralFunction}
\end{equation}

Since creation and annihilation operators are not Hermitian, they
cannot be used as $\sigma_{\mu}$ and $\sigma_{\nu}$ directly. A
trick consists in using a linear combination of the operators. For
each electron orbital, the Hermitian $X_{i\sigma}$ and $Y_{i\sigma}$
operators are defined from (\ref{eq:JordanWignerMap}) such that

\begin{equation}
\begin{array}{ccl}
X_{i\sigma} & \equiv & c_{i\sigma}+c_{i\sigma}^{\dagger}\\
\\
Y_{i\sigma} & \equiv & -i\left(c_{i\sigma}-c_{i\sigma}^{\dagger}\right).
\end{array}\label{eq:HermitianizedFermions}
\end{equation}
Note that $\left[X_{i\sigma},Y_{j\sigma'}\right]=i\delta_{ij}\delta_{\sigma\sigma'}Z_{i\sigma}$,
where $Z_{i\sigma}\equiv c_{i\sigma}^{\dagger}c_{i\sigma}-\frac{1}{2}$.
The elements of (\ref{eq:NambuGreensFunction}) can be computed from
the inverse transformation\begin{widetext}

\begin{equation}
\left(\begin{array}{c}
\left\langle c_{i\sigma}\left(\tau\right)c_{j\sigma'}^{\dagger}\left(0\right)\right\rangle \\
\left\langle c_{i\sigma}^{\dagger}\left(\tau\right)c_{j\sigma'}\left(0\right)\right\rangle \\
\left\langle c_{i\sigma}^{\phantom{}}\left(\tau\right)c_{j\sigma'}\left(0\right)\right\rangle \\
\left\langle c_{i\sigma}^{\dagger}\left(\tau\right)c_{j\sigma'}^{\dagger}\left(0\right)\right\rangle 
\end{array}\right)=\frac{1}{2}\left(\begin{array}{rrrr}
1 & 1 & i & -i\\
1 & 1 & -i & i\\
1 & -1 & i & i\\
1 & -1 & -i & -i
\end{array}\right)\left(\begin{array}{c}
\left\langle X_{i\sigma}\left(\tau\right)X_{j\sigma'}\left(0\right)\right\rangle \\
\left\langle Y_{i\sigma}\left(\tau\right)Y_{j\sigma'}\left(0\right)\right\rangle \\
\left\langle Y_{i\sigma}\left(\tau\right)X_{j\sigma'}\left(0\right)\right\rangle \\
\left\langle X_{i\sigma}\left(\tau\right)Y_{j\sigma'}\left(0\right)\right\rangle 
\end{array}\right)\label{eq:InverseJordanWigner}
\end{equation}
\end{widetext} Depending on the symmetries of the cluster Hamiltonian,
some terms in (\ref{eq:InverseJordanWigner}) may be zero at all time
and can be removed from the computation for speed-up or used to monitor
possible errors coming from noise or other sources.

\section{Conclusion}

We have outlined a method to compute different observables of the
FHM using a quantum computer. It synthesizes and builds mainly on
the work of \citep{Potthoff06,Senechal08,Riera12,Kaneko14,LasHeras15}.
Provided that the lattice can be divided into clusters (with $L_{c}$
spin-$\frac{1}{2}$ orbitals) which are coupled only with one-body
hopping terms, section \ref{sec:VariationalClusterApproximation}
reviewed how a variational principle for the grand canonical potential
of the model can be used to approximate the self-energy of the lattice
Hamiltonian and account for possible long-range ordering effects.
A similar construction where a functional would also integrate an
interaction across clusters \citep{Tong05} could also be considered.

The formalism to define a cluster was reviewed in section \ref{sec:ExampleSuperconductivity}
through the form of an example 2D lattice divided in $2\times2$ clusters
for which a few order parameters like antiferromagnetism and superconductivity
can be described and observable quantities computed. However, assuming
no spin, spatial or electron-hole symmetries in the cluster, up to
$4L_{c}^{2}$ variational terms can be defined. The nature of the
saddle-point problem that needs to be solved numerically is detailed
and the bottleneck is shown to be the diagonalization and the simulation
of the cluster which have to be solved for several variational parameters.

The scaling and solution methods for a given cluster are detailed
in section \ref{sec:EigenvalueProblem}. The memory scaling is known
to be very bad on classical computers as the dimension of the Hilbert
space of a cluster scales as $4^{L_{c}}$ in the number of orbitals.
A method which assumes some way of creating a Gibbs state at low temperature
on a quantum computer is presented. It is shown that there are $4L_{c}^{2}$
time correlation functions that need to be measured each round of
the saddle-point optimization problem. The Bravyi-Kitaev transformation
is known to significantly improve the scaling of classical algorithm
in the case where the number of electrons is conserved by the Hamiltonian
\citep{Seeley12} but a similar ansatz may also improve the method
presented in this paper (by dividing the Hilbert space in even/odd
occupation blocks for example).

This algorithm provides a novel way to simulate complex materials
at the electronic level and study new questions without knowing the
answer in advance. However some aspects could be improved. Notably,
it is not fully clear whether the transformation on the Gibbs state
be conditionally reversed after a measurement in such a way that the
state can be reused. The back-action of the correlation function measurement
may prevent the recycling of the Gibbs state. Also, it may be possible
to estimate the errors of the algorithm by simulating a known system
and comparing with analytical results (for example one could simulate
the well-known tight-binding model to benchmark the quantum algorithm).
Finally, it is possible that the method can be extended to simulate
non-equilibrium processes \citep{Nazarov09} by measuring the Keldysh
matrices $G^{R}$,$G^{A}$ and $G^{K}$. 
\begin{acknowledgments}
We are grateful to David Sénéchal for the very helpful discussion.
This work was supported by the European SCALEQIT program and Saarland
University.
\end{acknowledgments}

\appendix

\section{Numerical example on the 1D chain\label{sec:NumericalExample}}

The simplest experimental implementation of the variational procedure
on a quantum computer would correspond to solving a simple 1D tight-binding
chain. With a minimum cluster of $L_{c}=2$ sites (labeled ``$1$''
and ``$2$'') each with 2 electrons (spin-up and spin-down), a 5-qubit
quantum computer would be sufficient to extract the correlations functions
(\ref{eq:CorrelationFunction}). This section shows in detail how
the formalism of subsection \ref{sub:MeasuringCorrelationFunction}
can be used to compute the band structure and its occupation for the
1D chain at arbitrary $\text{\ensuremath{\mu}}$ and $T$. The simulation
was restricted only to a chemical variational potential $\mu'$ and
a simple pairing potential $\Delta'$ which is expected to be zero
in the case of one dimension.

\subsection{Finding the saddle-point of the self-energy functional\label{sub:NumericalSaddlePoint}}

First, the saddle point $\left(\begin{array}{c}
\mu'_{*}\\
\Delta'_{*}
\end{array}\right)$ of equation (\ref{eq:ExampleStationarityCondition}) must be found.
This is done through the following sequence:
\begin{enumerate}
\item \emph{Choose a point $\left(\begin{array}{c}
\mu'_{1}\\
\Delta'_{1}
\end{array}\right)$ and its neighbors $\left(\begin{array}{c}
\mu'_{1}\pm h\\
\Delta'_{1}
\end{array}\right)$ and $\left(\begin{array}{c}
\mu'_{1}\\
\Delta'_{1}\pm h
\end{array}\right)$ (with h a small parameter).}
\item \emph{On a quantum computer, measure the retarded Nambu Green's function
$\mathbf{\hat{G}'}\phantom{}^{R}\left(\tau,\mu',\Delta'\right)$ of
the cluster for the points of step 1 (as described in section \ref{sec:EigenvalueProblem}).}
\item \emph{Numerically compute the square of the gradient} (\ref{eq:ExampleStationarityCondition}).
\emph{If the modulus of the gradient is smaller than some threshold
$\epsilon_{\Omega}$, stop and assign $\left(\begin{array}{c}
\mu'_{*}\\
\Delta'_{*}
\end{array}\right)=\left(\begin{array}{c}
\mu'_{i}\\
\Delta'_{i}
\end{array}\right)$.}
\item \emph{Using a numerical Newton-Raphson method \citep{Benzi05}, pick
the next point $\left(\begin{array}{c}
\mu'_{i+1}\\
\Delta'_{i+1}
\end{array}\right)$ and loop over to step 1.}
\end{enumerate}
Once the saddle-point is known, $\mathbf{\hat{G}'}\phantom{}^{R}\left(\tau,\mu'_{*},\Delta'_{*}\right)$
is measured and properties like the spectral density of the lattice
can be approximated.

\subsection{Measuring and calculating the retarded Green's function of the cluster\label{sub:Solving1DCluster}}

The retarded Nambu Green's function is measured on a discrete time
domain $\tau_{n}=n\Delta\tau$ where n is an integer between 0 and
$n_{\mathrm{max}}$ and $\Delta\tau$ is a small time interval ($n_{\mathrm{max}}=2000$
and $\Delta\tau=0.05$ in this example) such that $\tau_{\mathrm{max}}=n_{\mathrm{max}}\Delta\tau$.
The matrix form of $\mathbf{\hat{G}'}\phantom{}^{R}$ clearly shows
that the number of correlation functions $\left\langle c_{\mu}\left(\tau\right)c_{\nu}^{\dagger}\left(0\right)\right\rangle $
scales as $4L_{c}^{2}$:\begin{widetext}
\begin{equation}
\mathbf{\hat{G}'}\phantom{}^{R}\left(\tau_{n}\right)=-i\theta\left(\tau_{n}\right)\left(\begin{array}{cccc}
\left\langle c_{1\uparrow}\left(\tau_{n}\right)c_{1\uparrow}^{\dagger}\left(0\right)\right\rangle  & \left\langle c_{1\uparrow}\left(\tau_{n}\right)c_{2\uparrow}^{\dagger}\left(0\right)\right\rangle  & \left\langle c_{1\uparrow}^{\phantom{}}\left(\tau_{n}\right)c_{1\downarrow}\left(0\right)\right\rangle  & \left\langle c_{1\uparrow}^{\phantom{}}\left(\tau_{n}\right)c_{2\downarrow}\left(0\right)\right\rangle \\
\left\langle c_{2\uparrow}\left(\tau_{n}\right)c_{1\uparrow}^{\dagger}\left(0\right)\right\rangle  & \left\langle c_{2\uparrow}\left(\tau_{n}\right)c_{2\uparrow}^{\dagger}\left(0\right)\right\rangle  & \left\langle c_{2\uparrow}^{\phantom{}}\left(\tau_{n}\right)c_{1\downarrow}\left(0\right)\right\rangle  & \left\langle c_{2\uparrow}^{\phantom{}}\left(\tau_{n}\right)c_{2\downarrow}\left(0\right)\right\rangle \\
\left\langle c_{1\downarrow}^{\dagger}\left(\tau_{n}\right)c_{1\uparrow}^{\dagger}\left(0\right)\right\rangle  & \left\langle c_{1\downarrow}^{\dagger}\left(\tau_{n}\right)c_{2\uparrow}^{\dagger}\left(0\right)\right\rangle  & \left\langle c_{1\downarrow}^{\dagger}\left(\tau_{n}\right)c_{1\downarrow}\left(0\right)\right\rangle  & \left\langle c_{1\downarrow}^{\dagger}\left(\tau_{n}\right)c_{2\downarrow}\left(0\right)\right\rangle \\
\left\langle c_{2\downarrow}^{\dagger}\left(\tau_{n}\right)c_{1\uparrow}^{\dagger}\left(0\right)\right\rangle  & \left\langle c_{2\downarrow}^{\dagger}\left(\tau_{n}\right)c_{2\uparrow}^{\dagger}\left(0\right)\right\rangle  & \left\langle c_{2\downarrow}^{\dagger}\left(\tau_{n}\right)c_{1\downarrow}\left(0\right)\right\rangle  & \left\langle c_{2\downarrow}^{\dagger}\left(\tau_{n}\right)c_{2\downarrow}\left(0\right)\right\rangle 
\end{array}\right).\label{eq:RetardedRealTimeGreensFunction1D}
\end{equation}
 \end{widetext} It is then Fourier transformed on a discrete frequency
domain $\omega_{m}=m\Delta\omega$ between $-\omega_{\mathrm{max}}$
and $\omega_{\mathrm{max}}$ chosen such that $\omega_{\mathrm{max}}=\frac{1}{2\Delta\tau}$
and $\Delta\omega=\frac{1}{2\tau_{\mathrm{max}}}$ :
\begin{equation}
\mathbf{\hat{G}'}\phantom{}^{R}\left(\omega_{m}\right)=\frac{\Delta\tau}{2\pi}\sum_{n=0}^{n_{\mathrm{max}}}e^{-i\omega_{m}\tau_{n}}\mathbf{\hat{G}'}\phantom{}^{R}\left(\tau_{n}\right).\label{eq:RetardedFrequencyGreensFunction1D}
\end{equation}
The numerical $\mathbf{\hat{G}'}\phantom{}^{R}\left(\omega\right)$
can then be used to compute the lattice-perturbed Green's function
$\mathbf{\mathcal{\hat{G}}}\left(\mathbf{\mathbf{\tilde{k}},\omega}\right)$
(see equation (\ref{eq:LatticePerturbedGreensFunction})) and various
properties of the lattice as detailed in subsection \ref{sub:CalculationOfObservables}.
The exact mapping of (\ref{eq:RetardedRealTimeGreensFunction1D})
on the quantum computer is done through the Jordan-Wigner transformation

\begin{equation}
\begin{array}{ccl}
c_{1\uparrow}^{\dagger} & = & \mathbb{I}\otimes\mathbb{I}\otimes\mathbb{I}\otimes\sigma_{+}\\
\\
c_{2\uparrow}^{\dagger} & = & \mathbb{I}\otimes\mathbb{I}\otimes\sigma_{+}\otimes\sigma_{z}\\
\\
c_{1\downarrow}^{\dagger} & = & \mathbb{I}\otimes\sigma_{+}\otimes\sigma_{z}\otimes\sigma_{z}\\
\\
c_{2\downarrow}^{\dagger} & = & \sigma_{+}\otimes\sigma_{z}\otimes\sigma_{z}\otimes\sigma_{z}.
\end{array}\label{eq:JordanWignerMap1D}
\end{equation}
\medskip{}
Using this transformation, all component of the Hamiltonian $\mathcal{H}'$
of the cluster (\ref{eq:HamiltonianOneCluster}) are mapped to a 4-qubit
Hilbert space: \begin{widetext}

\begin{equation}
\begin{array}{ccl}
\mathcal{H}_{\mathrm{FH}} & = & -t\left(c_{1\uparrow}^{\dagger}c_{2\uparrow}+c_{2\uparrow}^{\dagger}c_{1\uparrow}+c_{1\downarrow}^{\dagger}c_{2\downarrow}+c_{2\downarrow}^{\dagger}c_{\downarrow\uparrow}\right)-U\left(n_{1\uparrow}n_{1\downarrow}+n_{2\uparrow}n_{2\downarrow}\right)\\
\\
 & = & -t\left(\mathbb{I}\otimes\mathbb{I}\otimes\left(\sigma_{-}\otimes\sigma_{+}+\sigma_{+}\otimes\sigma_{-}\right)+\left(\sigma_{-}\otimes\sigma_{+}+\sigma_{+}\otimes\sigma_{-}\right)\otimes\mathbb{I}\otimes\mathbb{I}\right)\\
\\
 &  & -U\left(\mathbb{I}\otimes\sigma_{n}\otimes\mathbb{I}\otimes\sigma_{n}+\sigma_{n}\otimes\mathbb{I}\otimes\sigma_{n}\otimes\mathbb{I}\right),
\end{array}\label{eq:FermiHubbardHamiltonianCluster1D}
\end{equation}

\begin{equation}
\begin{array}{ccl}
\mathcal{H}_{\mathrm{pair}} & = & \Delta'\left(c_{1\uparrow}^{\dagger}c_{1\downarrow}^{\dagger}+c_{1\downarrow}c_{1\uparrow}+c_{2\uparrow}^{\dagger}c_{2\downarrow}^{\dagger}+c_{2\downarrow}c_{2\uparrow}\right)\\
\\
 & = & \Delta'\left(\mathbb{I}\otimes\left(\sigma_{+}\otimes\sigma_{z}\otimes\sigma_{+}+\sigma_{-}\otimes\sigma_{z}\otimes\sigma_{-}\right)+\left(\sigma_{+}\otimes\sigma_{z}\otimes\sigma_{+}+\sigma_{-}\otimes\sigma_{z}\otimes\sigma_{-}\right)\otimes\mathbb{I}\right),
\end{array}\label{eq:PairingHamiltonian1D}
\end{equation}

\begin{equation}
\begin{array}{ccl}
\mathcal{H}_{\mathrm{local}} & = & \mu'\left(n_{1\uparrow}+n_{2\uparrow}+n_{1\downarrow}+n_{2\downarrow}\right)\\
\\
 & = & \mu'\left(\mathbb{I}\otimes\mathbb{I}\otimes\mathbb{I}\otimes\sigma_{n}+\mathbb{I}\otimes\mathbb{I}\otimes\sigma_{n}\otimes\mathbb{I}+\mathbb{I}\otimes\sigma_{n}\otimes\mathbb{I}\otimes\mathbb{I}+\sigma_{n}\otimes\mathbb{I}\otimes\mathbb{I}\otimes\mathbb{I}\right).
\end{array}\label{eq:VariationChemicalPotential1D}
\end{equation}
\end{widetext} 

It can be noticed that the standard Fermi-Hubbard term requires gates
between two qubits, the variational chemical potential can be implemented
with single qubit gates but the pairing terms need operations over
several qubits to maintain the statistics of the fermions. The perturbation
matrix (\ref{eq:ClusterPerturbationPotential}) is given explicitly
by
\begin{equation}
\hat{\mathbf{V}}\left(\mathbf{\tilde{k}}\right)=\left(\begin{array}{cccc}
-\mu+\mu' & \epsilon\left(\tilde{k}\right)+t & -\Delta' & 0\\
\epsilon^{*}\left(\tilde{k}\right)+t & -\mu+\mu' & 0 & -\Delta'\\
-\Delta' & 0 & \mu-\mu' & -\epsilon\left(\tilde{k}\right)-t\\
0 & -\Delta' & -\epsilon^{*}\left(\tilde{k}\right)-t & \mu-\mu'
\end{array}\right).\label{eq:ClusterPerturbationPotential1D}
\end{equation}
Finally the operators that are applied in the phase estimation part
of the algorithm and are required in the reconstruction of (\ref{eq:RetardedRealTimeGreensFunction1D})
are given by the following transformations:\begin{widetext}

\begin{equation}
\begin{array}{rcrclcrcrcl}
X_{1\uparrow} & = & c_{1\uparrow}+c_{1\uparrow}^{\dagger} & = & \frac{1}{2}\mathbb{I}\otimes\mathbb{I}\otimes\mathbb{I}\otimes\sigma_{x}, &  & Y_{1\uparrow} & = & -i\left(c_{1\uparrow}-c_{1\uparrow}^{\dagger}\right) & = & \frac{1}{2}\mathbb{I}\otimes\mathbb{I}\otimes\mathbb{I}\otimes\sigma_{y},\\
\\
X_{2\uparrow} & = & c_{2\uparrow}+c_{2\uparrow}^{\dagger} & = & \frac{1}{2}\mathbb{I}\otimes\mathbb{I}\otimes\sigma_{x}\otimes\sigma_{z}, &  & Y_{2\uparrow} & = & -i\left(c_{2\uparrow}-c_{2\uparrow}^{\dagger}\right) & = & \frac{1}{2}\mathbb{I}\otimes\mathbb{I}\otimes\sigma_{y}\otimes\sigma_{z},\\
\\
X_{1\downarrow} & = & c_{1\downarrow}+c_{1\downarrow}^{\dagger} & = & \frac{1}{2}\mathbb{I}\otimes\sigma_{x}\otimes\sigma_{z}\otimes\sigma_{z}, &  & Y_{1\downarrow} & = & -i\left(c_{1\downarrow}-c_{1\downarrow}^{\dagger}\right) & = & \frac{1}{2}\mathbb{I}\otimes\sigma_{y}\otimes\sigma_{z}\otimes\sigma_{z},\\
\\
X_{2\downarrow} & = & c_{2\downarrow}+c_{2\downarrow}^{\dagger} & = & \frac{1}{2}\sigma_{x}\otimes\sigma_{z}\otimes\sigma_{z}\otimes\sigma_{z}, &  & Y_{2\downarrow} & = & -i\left(c_{2\downarrow}-c_{2\downarrow}^{\dagger}\right) & = & \frac{1}{2}\sigma_{y}\otimes\sigma_{z}\otimes\sigma_{z}\otimes\sigma_{z}.
\end{array}\label{eq:HermitianizedFermions1D}
\end{equation}
\end{widetext}

The procedure highlighted in subsection \ref{sub:SuperlatticeClusters}
is then followed to compute the CPT Green's function and the desired
properties of the system.

\subsection{Simple tight-binding model}

The tight-binding model $U=0$ is investigated using the methods of
this paper. The goal is to show that the method can accurately simulate
well known simple models through the intermediate results it produces. 

\begin{figure}
\begin{centering}
\includegraphics[width=3.375in]{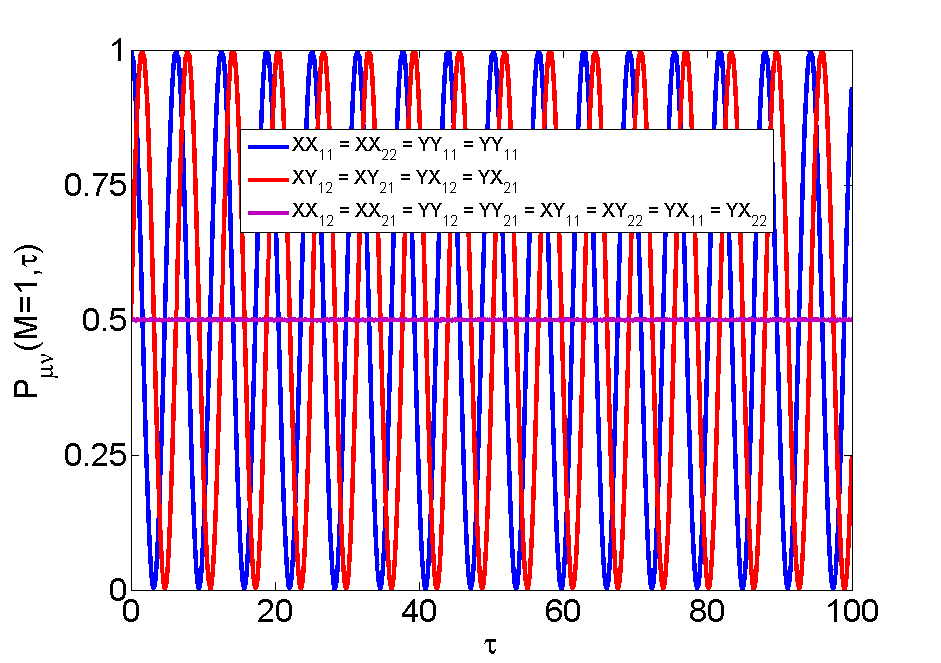}\caption{Measured probabilities for different $X_{\mu}$ and $Y_{\mu}$ at
different times. In this case the cluster parameters are $L_{c}=2$,
$t=1$, $U=\Delta'=\mu'=0$ and $T=0.1$.\label{fig:MeasuredProbabilities}}

\par\end{centering}

\end{figure}

In figure \ref{fig:MeasuredProbabilities}, the measured value of
$P_{\mu\nu}\left(\mathcal{M}=1,\tau\right)$ is shown for the simplest
case of a 2-site tight-binding cluster. In this case the model generates
simple oscillations as no decoherence is included.

\begin{figure}
\begin{centering}
\includegraphics[width=3.375in]{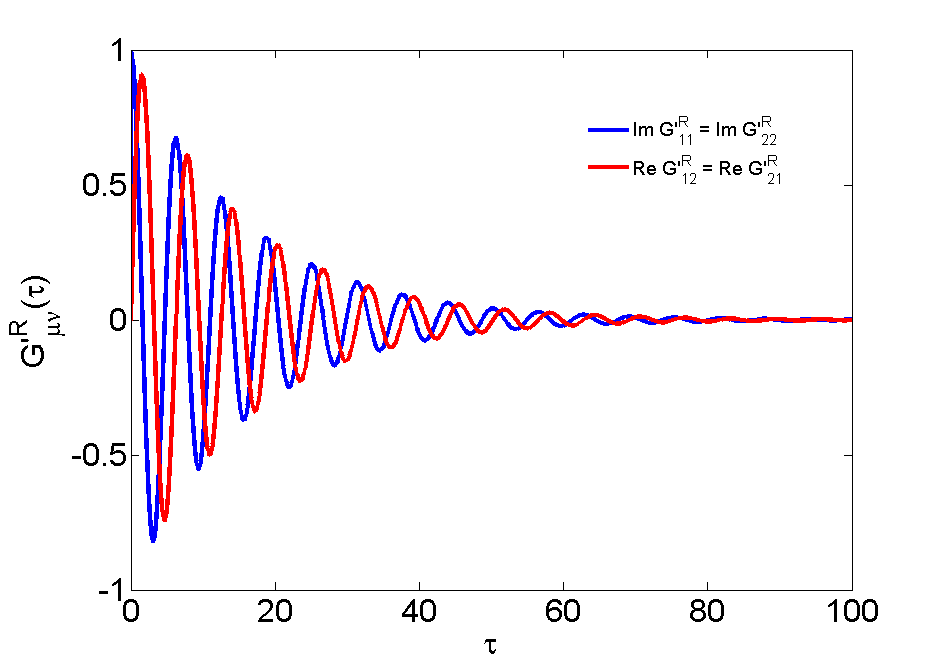}
\par\end{centering}

\caption{Non-zero correlation functions computed from the results of figure
\ref{fig:MeasuredProbabilities}. The function was regularized with
a $e^{-\text{\ensuremath{\eta\tau}}}$ term to remove the fast oscillations
of the Fourier transform arising from the finiteness of the time domain,
$\eta=\frac{\pi}{50}$ was used in this case.\label{fig:MeasuredGreensFunctions}}
\end{figure}

In figure \ref{fig:MeasuredGreensFunctions}, the Green's functions
$G'_{\mu\nu}\phantom{}^{R}\left(\tau\right)$ computed from equation
(\ref{eq:InverseJordanWigner}) are shown. Notice that the time-dependent
Green's functions were regularized with an decaying exponential $e^{-\text{\ensuremath{\eta\tau}}}$
in order to remove the fast oscillations coming from the convolution
of the frequency-dependent Green's function with the $\mathrm{sinc}\left(\frac{\omega\tau_{\mathrm{max}}}{2\pi}\right)$
term involved in finite time measurements. This regularizing term
is not decoherence, but it could model a uniform depolarizing rate
$\eta$ in the quantum processor. This rate would actually contribute
to the width of the frequency-dependent Green's function.

\begin{figure}
\begin{centering}
\includegraphics[width=3.375in]{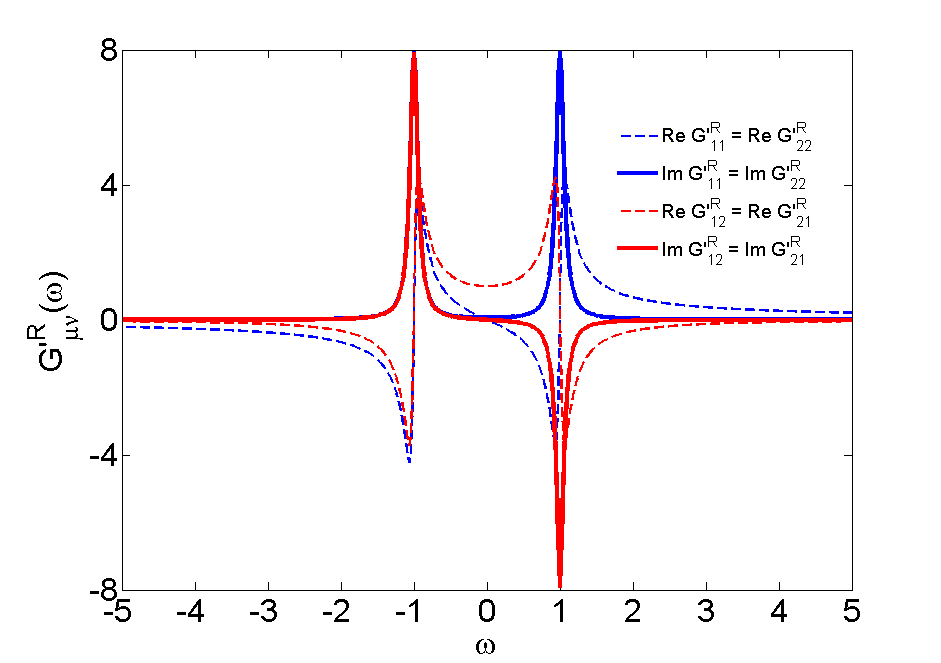}
\par\end{centering}

\caption{Real and imaginary parts of the frequency-dependent Green's functions
arising from the correlation functions measured in figure \ref{fig:MeasuredGreensFunctions}.\label{fig:MeasuredFrequencyGF}}

\end{figure}

In figure \ref{fig:MeasuredFrequencyGF}, the Fourier transformed
$G'_{\mu\nu}\phantom{}^{R}\left(\omega\right)$ are shown for the
simple tight-binding cluster. Only two peaks are present and their
width is determined by $\eta$ and the time domain used to measure
the correlation functions.

\begin{figure}
\begin{centering}
\includegraphics[width=3.375in]{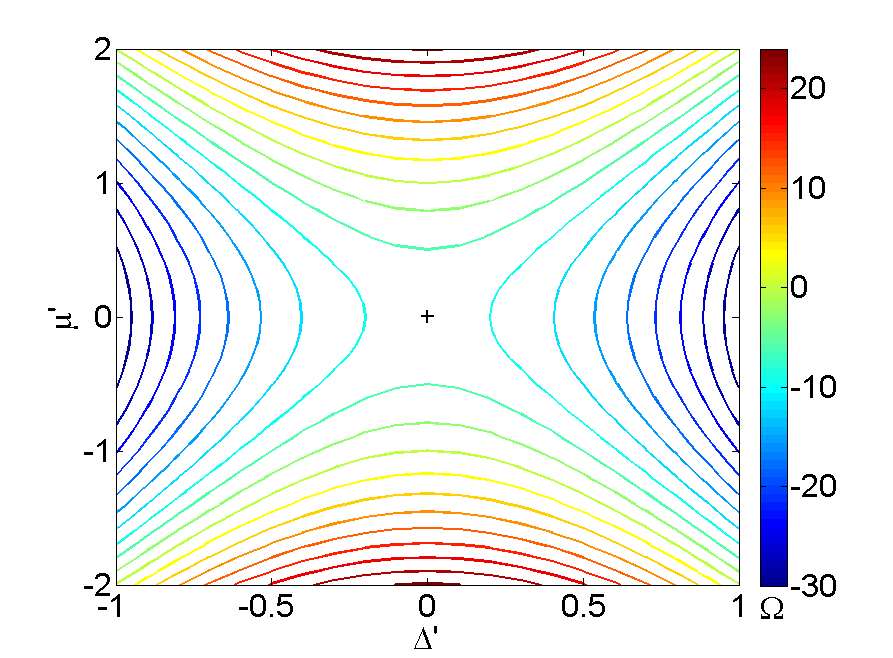}
\par\end{centering}

\caption{Potthoff functional $\Omega$ for different variational parameters
$\mu'$ and $\Delta'$ of a cluster of size $L_{c}=2$ with parameters
$t=1$, $U=0$, $\mu=0$ and $T=1$. The cross marks the saddle point
at $\left(\protect\begin{array}{c}
\mu'_{*}\protect\\
\Delta'_{*}
\protect\end{array}\right)=\left(\protect\begin{array}{c}
0.0046\protect\\
0
\protect\end{array}\right)$.\label{fig:PotthoffFunctional}}

\end{figure}

Figure \ref{fig:PotthoffFunctional} shows an example of the Potthoff
functional $\Omega\left(\mu',\Delta'\right)$ and its saddle point
for a small 1D cluster.  As expected for this simple model, the saddle
point is almost at the origin, the small deviation comes from the
low finite temperature. At the saddle point, the average occupation
of each state is $\left\langle n\right\rangle =0.5$ as is expected.
At the saddle-point the spectral density of the full lattice can be
computed.

\begin{figure}
\begin{centering}
\includegraphics[width=3.375in]{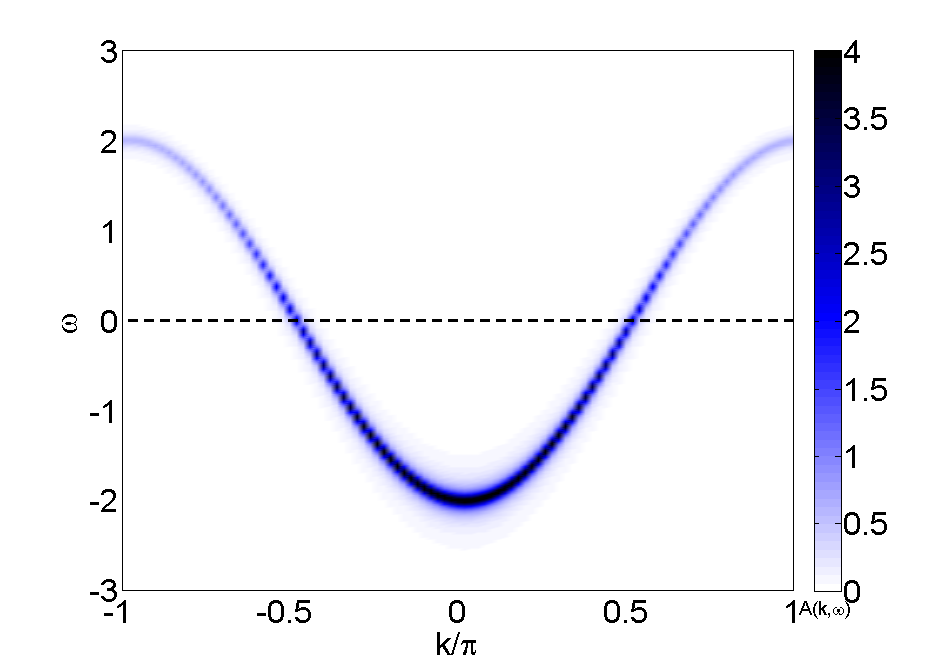}
\par\end{centering}

\caption{Electron momentum-frequency distribution $A\left(k,\omega\right)$
for a lattice with parameters $t=1$, $U=0$, $\mu=0$ and $T=1$.
The cluster used had $L_{c}=2$ site and the saddle-point is the same
as in figure \ref{fig:PotthoffFunctional}. The dashed line is at
the chemical potential.\label{fig:SpectralDensity} }
\end{figure}

Figure \ref{fig:SpectralDensity} shows the spectral density $A\left(k,\omega\right)$computed
from equation (\ref{eq:QuasiparticleSpectrum}) for 50 clusters of
size $L_{c}=2$ in a simple tight binding model at relatively high
temperature $T=1$. The cosine band is fill above the Fermi level
because of the high temperature.

\begin{figure}
\begin{centering}
\includegraphics[width=3.375in]{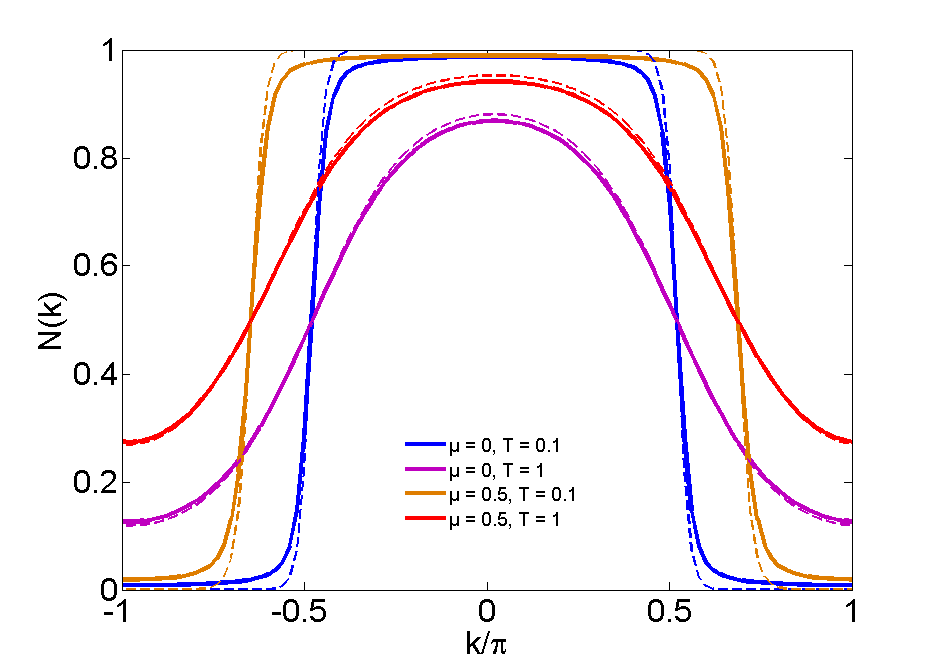}
\par\end{centering}

\caption{Electron momentum distribution $N\left(k\right)$ for different chemical
potentials $\mu$ and temperatures $T$ with $U=0$. The solid lines
are the results from the numerical simulation of the quantum algorithm
using time steps of size $d\tau=0.02$ up to $\tau_{\mathrm{max}}=200$
while the dashed lines come from an imaginary frequency summation.\label{fig:DensityofStates}}
\end{figure}

Figure \ref{fig:DensityofStates} shows that the simulation yields
the expected physics of the tight-binding model at finite temperature.
The ground state is indeed a 1D Fermi sea in the electronic momentum
distribution (\ref{eq:MomentumDistribution}) whose width is increased
with the chemical potential and broadened by increased temperature.
The loss of accuracy in the simulation is attributed to the sampling
method and the accuracy of the Fermi distribution on the discrete
frequency domain computed from the measured time series.

\begin{figure}
\begin{centering}
\includegraphics[width=3.375in]{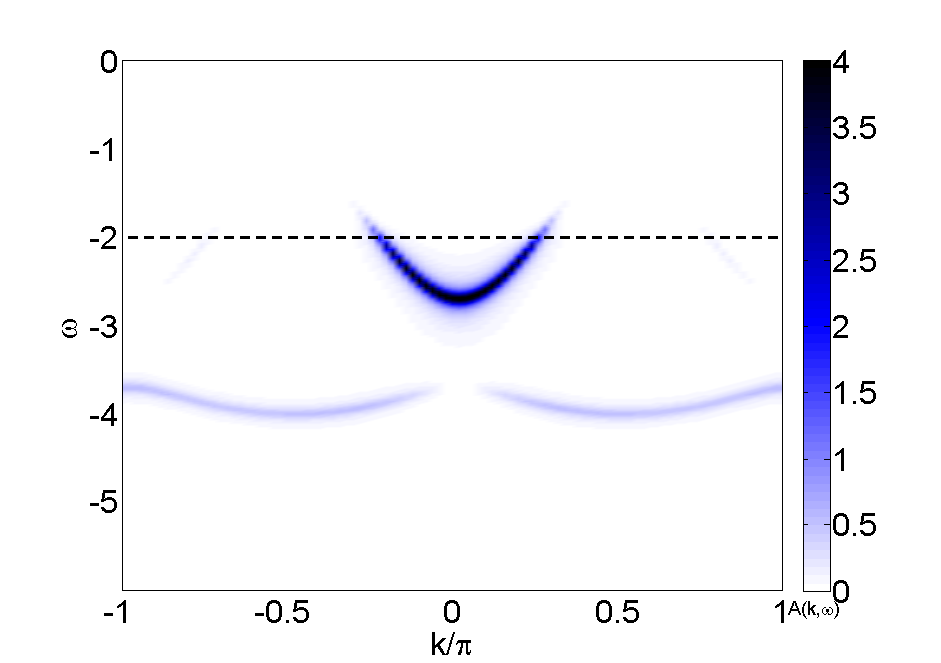}
\par\end{centering}

\caption{Electron momentum-frequency distribution $A\left(k,\omega\right)$
for a lattice with parameters $t=1$ , $U=4$ , $\mu=-2$ and $T=0.1$
. The cluster used had $L_{c}=2$ site and the saddle-point is at
$\left(\protect\begin{array}{c}
\mu'_{*}\protect\\
\Delta'_{*}
\protect\end{array}\right)=\left(\protect\begin{array}{c}
-2\protect\\
0
\protect\end{array}\right)$. The dashed line is at the chemical potential.\label{fig:SpectralDensityU}}
\end{figure}

Finally figure \ref{fig:SpectralDensityU} shows the spectral density
$A\left(k,\omega\right)$ computed from equation \ref{eq:QuasiparticleSpectrum}
for a cluster of size $L_{c}=2$ in an attractive Hubbard chain $U=4$
at low temperature $T=0.1$. The band is highly distorted by the interaction
and the ground state is no longer a $k=0$ state.

Extending these calculation for more complicated model is an easy
task. A simple 2D model with a superconducting phase transition would
require 4 sites and 8 electrons, so a 9-qubit quantum computer would
be required to measure $\mathbf{\hat{G}'}\phantom{}^{R}\left(\tau\right)$
in this case.\emph{  }It appears that the number of time points
that need to be measured may become an issue as the systems become
more complex. It would be interesting to know if there exist sampling
methods as efficient as imaginary frequency summation methods \citep{Ozaki07}
where only $\approx100$ points need to be measured in order to achieve
a high numerical accuracy in the computation of the Fermi function
even for complicated electronic structures. For example, a cost function
over several models could be used to extract the Green's function
using fewer measurements. Alternatively, measuring forward finite
difference time derivatives close to $\tau=0$ to get the coefficients
of the moment expansion of equation (\ref{eq:LatticePerturbedGreensFunction})
could also work. Indeed, the correlation functions (\ref{eq:CorrelationFunction})
can be rewritten as
\begin{equation}
C_{\mu\nu}\left(\tau\right)=\sum_{s=0}^{\infty}\frac{\tau^{s}}{s!}C_{\mu\nu}^{\left(s\right)}\label{eq:CorrelationFunctionMomentExpansion}
\end{equation}
where the moments are given by
\begin{equation}
\begin{array}{rcl}
C_{\mu\nu}^{\left(s\right)} & = & \left(-i\right)^{s}\sum_{m}\sum_{n}A_{\mu\nu}^{mn}\left(E_{m}-E_{n}\right)^{s}\\
\\
 & = & \underset{\tau\rightarrow0}{\mathrm{lim}}\frac{d^{s}}{d\tau^{s}}C_{\mu\nu}\left(\tau\right)\\
\\
 & = & \left(\Delta\tau\right)^{-s}\sum_{r=0}^{s}\left(-1\right)^{r}\left(\begin{array}{c}
s\\
r
\end{array}\right)C_{\mu\nu}\left(\left(s-r\right)\Delta\tau\right)\\
 &  & +O\left(\Delta\tau\right)
\end{array}\label{eq:CorrelationFunctionMoments}
\end{equation}
which could be approximated experimentally by forward finite differences
(higher order finite differences could also be used).

\section{Preparation of a Gibbs state\label{sec:GibbsStatePreparation}}

A digital method to prepare Gibbs states in a quantum computer is
reviewed and shown adequate for a variational solver. The goal is
the make this document self-contained in the sense that action of
the quantum computer can be fully defined.

\begin{figure}
\begin{centering}
\includegraphics[width=3.375in]{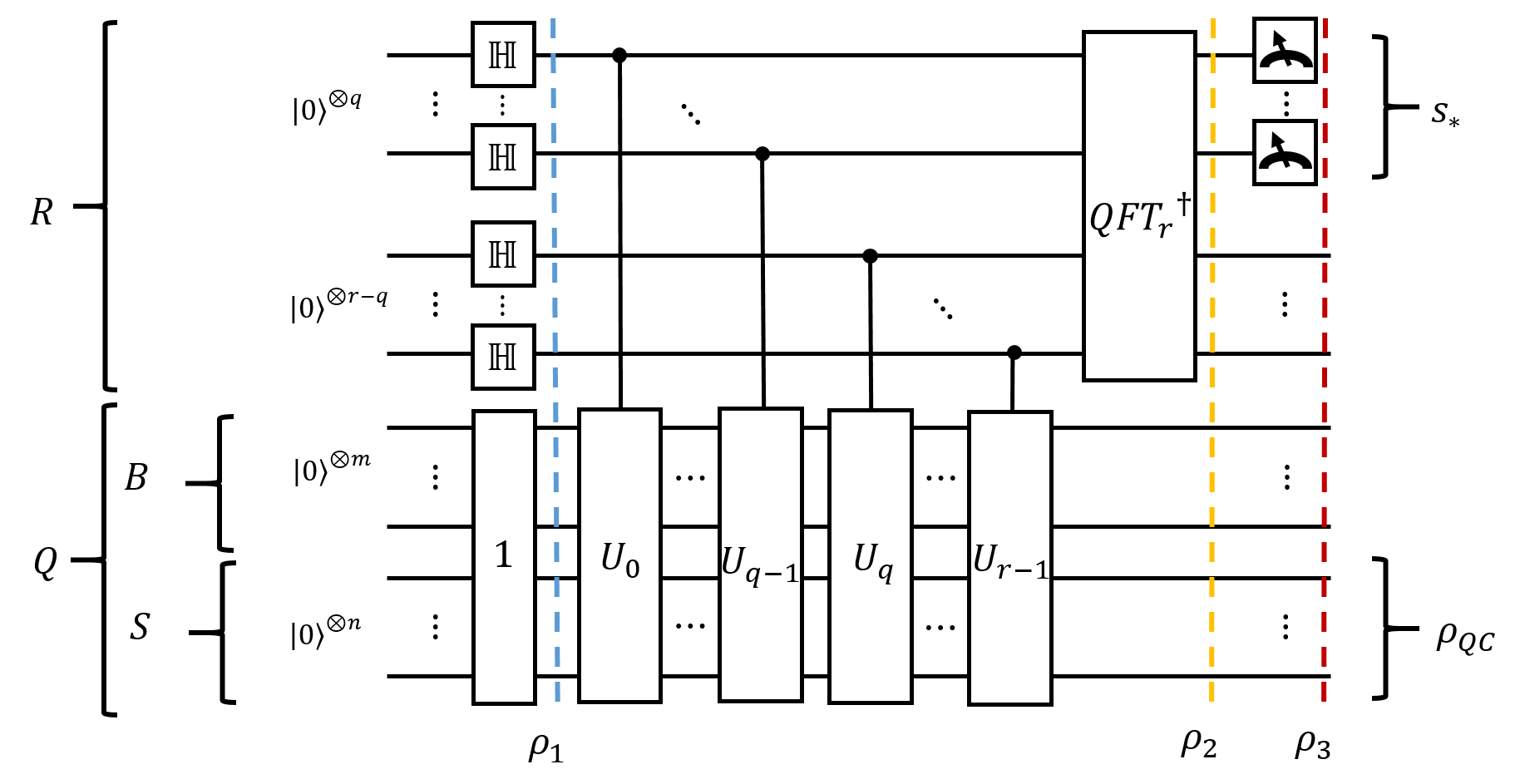}
\par\end{centering}

\caption{Detailed circuit to prepare an approximate Gibbs state $\rho_{QC}\approx\rho_{\mathrm{Gibbs}}$
following \citep{Riera12}. The simulated inverse temperature $\beta$
is related to the measurement of $s_{*}$ by equation (\ref{eq:SimulatedInverseTemperature}).
The initial state of $R$ and $Q$ is taken to be the zero state $\left|0\right\rangle ^{\otimes\left(q+m+n\right)}$,
then the Hadamard gate $\mathbb{H}^{\otimes q}$ is applied on $R$
and $Q$ is transformed (non-unitarily) to the fully mixed state $\frac{1}{2^{m+n}}\mathbb{I}^{\otimes\left(m+n\right)}$.
Then $q$ controlled-$U$ operations are applied, where the notation
$U_{\tau}=U^{2^{\tau}}$ and $U=e^{-i\frac{\mathcal{H}_{0}}{\left\Vert \mathcal{H}_{0}\right\Vert _{\infty}}}$
with $\mathcal{H}_{0}=\mathcal{H}'+\mathcal{H}_{B}$. An inverse quantum
Fourier transform is applied on register $R$ and the string $s_{*}$
is read from the first $q$ qubits. Register $S$ is then left in
a simulated Gibbs state $\rho_{QC}^{S}$.\label{fig:GibbsCircuit}}
\end{figure}

Here is the summary of the method, as given in \citep{Riera12}, to
prepare the Gibbs state required to simulate the correlation function
of the cluster. In addition to the simulated system Hamiltonian $\mathcal{H}'$,
a bath Hamiltonian $\mathcal{H}_{B}$ is required such that the total
uncoupled system is
\begin{equation}
\mathcal{H}_{0}=\mathcal{H}'+\mathcal{H}_{B}\label{eq:HamiltonianUncoupledSystemBath}
\end{equation}
with eigenvalues $\left\{ E_{k}^{(0)}\right\} $ and energy eigenvectors
$\left\{ \left|E_{k}^{(0)}\right\rangle \right\} $. The bath (first
part of the register $Q$ in figure \ref{fig:FullCircuit}) is assumed
to be a collection of $m$ uncoupled spin-$\frac{1}{2}$ with energy
splitting $\eta$:

\begin{equation}
\mathcal{H}_{B}=\frac{\eta}{2}\sum_{j=1}^{m}\left(\mathbb{I}_{j}+\sigma_{zj}\right).\label{eq:HamiltonianBath}
\end{equation}
A small interaction $\mathcal{V}$ is allowed such that the total
coupled system Hamiltonian is

\begin{equation}
\mathcal{H}_{tot}=\mathcal{H}_{0}+\mathcal{V}\label{eq:HamiltonianCoupledSystemBath}
\end{equation}
with eigenvalues $\left\{ E_{k}\right\} $ and energy eigenvectors
$\left\{ \left|E_{k}\right\rangle \right\} $. The procedure is the
following (see figure \ref{fig:GibbsCircuit})
\begin{enumerate}
\item \emph{Initialization. $r$ Hadamard gates $\mathbb{H}$ are applied
on the qubits of register $R$ and the register $Q$ is relaxed in
the fully mixed state of (\ref{eq:HamiltonianCoupledSystemBath})
such that
\begin{equation}
\rho_{1}=\frac{1}{d}\sum_{s,s'=0}^{2^{r}-1}\left|s\right\rangle \left\langle s'\right|\otimes\sum_{k=1}^{d}\left|E_{k}\right\rangle \left\langle E_{k}\right|\label{eq:GibbsStatePrepStep1}
\end{equation}
where $d=2^{m+2L_{c}}$ is the total dimension of the system plus
bath. This is equivalent to preparing the coupled system + bath at
infinite temperature.}
\item \emph{Partial quantum phase estimation. $r$ controlled-$U$ operation
are followed by an inverse Fourier transform on $R$. Note that $U=e^{-i\frac{\mathcal{H}_{0}}{\left\Vert \mathcal{H}_{0}\right\Vert _{\infty}}}$,
where $\mathcal{H}_{0}$ is the uncoupled Hamiltonian (\ref{eq:HamiltonianUncoupledSystemBath}).
After this phase estimation part, the state in the computer is
\begin{equation}
\rho_{2}=\frac{1}{d}\sum_{s,s'=0}^{2^{r}-1}\sum_{k=1}^{d}\alpha_{s}\left(\varphi_{k}\right)\alpha_{s'}^{*}\left(\varphi_{k}\right)\left|s\right\rangle \left\langle s'\right|\otimes\left|E_{k}\right\rangle \left\langle E_{k}\right|\label{eq:GibbsStatePrepStep2}
\end{equation}
where $\varphi_{k}\equiv\frac{E_{k}}{\left\Vert \mathcal{H}_{tot}\right\Vert _{\infty}}$
and
\begin{equation}
\alpha_{s}\left(\varphi\right)\equiv\frac{1}{2^{r}}\frac{1-e^{2\pi i\left(2^{r}\varphi-s\right)}}{1-e^{2\pi i\left(\varphi-2^{-r}s\right)}}\label{eq:PhaseDefStep2}
\end{equation}
The controlled evolution of the full system dephases different distributions
of eigenvalues contained in the fully mixed state.}
\item \emph{Measurement. The first $q$ qubits of $R$ are measured. A binary
string $s_{*}$ (length $q$) is obtained
\begin{equation}
\rho_{3}\propto\sum_{s,s'=s_{*}\Delta_{\mathrm{rect}*}}^{\left(s_{*}+1\right)\Delta_{\mathrm{rect}*}}\sum_{k=1}^{d}\alpha_{s}\left(\varphi_{k}\right)\alpha_{s'}^{*}\left(\varphi_{k}\right)\left|s\right\rangle \left\langle s'\right|\otimes\left|E_{k}\right\rangle \left\langle E_{k}\right|\label{eq:GibbsStatePrepStep3}
\end{equation}
where $\Delta_{\mathrm{rect}*}\equiv2^{r-q}$ is the number of states
of the ancillary register $R$ compatible with the measurement. 
The width of the rectangular state that is prepared is determined
by $\Delta_{\mathrm{rect}}=\left\Vert \mathcal{H}_{tot}\right\Vert _{\infty}2^{-r}\Delta_{\mathrm{rect}*}$.
The energy of the rectangular state is $E=\left\Vert \mathcal{H}_{tot}\right\Vert _{\infty}2^{-q}s_{*}$.
The inverse temperature $\beta$ is determined by $E$ and $\Delta_{\mathrm{rect}}$.
The final state in the register $Q$ is now
\begin{equation}
\begin{array}{rcl}
\rho_{QC} & \equiv & \textrm{Tr}_{R}\rho_{3}\\
\\
 & \propto & \sum_{k=1}^{d}\left(\sum_{s=s_{*}\Delta_{\mathrm{rect}}}^{\left(s_{*}+1\right)\Delta_{\mathrm{rect}}}\left|\alpha_{s}\left(\varphi_{k}\right)\right|^{2}\right)\left|E_{k}\right\rangle \left\langle E_{k}\right|.
\end{array}\label{eq:StateRegisterQ}
\end{equation}
One of the rectangular states contained in the initial fully mixed
state is selected upon measurement. For appropriately chosen parameters,
the state in register $S$ is approximately a Gibbs state of the cluster
Hamiltonian.}
\end{enumerate}
The algorithm outputs a reduced state $\rho_{QC}^{S}=\textrm{Tr}_{B}\rho_{QC}\approx\rho_{\textrm{Gibbs}}^{S}=\frac{e^{-\beta\mathcal{H}'}}{\textrm{Tr}e^{-\beta\mathcal{H}'}}$
in the channel $S$, where $\beta=\frac{1}{T}$ is the inverse temperature.
Assuming a bath of the form (\ref{eq:HamiltonianBath}) with energy
scale $\eta=\sqrt{\frac{\lambda}{m}}\left\Vert \mathcal{H}'\right\Vert _{\infty}$,
the ``$\approx$'' really implies the following condition

\begin{equation}
\begin{array}{rcl}
\mathcal{D}\left(\rho_{QC}^{S},\rho_{\textrm{Gibbs}}^{S}\right) & \leq & \left(1+\frac{\ln\left(2^{r-q}\right)}{\pi^{2}}\right)\frac{e^{\frac{2}{\lambda}+\beta\left\Vert \mathcal{H}'\right\Vert _{\infty}+\frac{\lambda\left\Vert \mathcal{H}'\right\Vert _{\infty}^{2}\beta^{2}}{8}}}{2^{r-q-2}}\\
\\
 &  & +\frac{1}{2}\left(e^{\frac{2}{\lambda}}-1\right)+C
\end{array}\label{eq:StatePreparationErrorBound}
\end{equation}
where $\mathcal{D}\left(\cdot,\cdot\right)$ is the trace distance
and $C$ is a constant exponentially small in $m$. \emph{} The effective
inverse temperature is in the interval $\left[\beta-\delta\beta,\beta+\delta\beta\right]$
with 
\begin{equation}
\beta=\frac{4}{\eta}\left(\frac{1}{2}-2^{-q}s_{*}\left(1+\frac{\left\Vert \mathcal{H}'\right\Vert _{\infty}}{\left\Vert \mathcal{H}_{B}\right\Vert _{\infty}}\right)\right).\label{eq:SimulatedInverseTemperature}
\end{equation}
Since $s_{*}\in\left[0,2^{q}-1\right]$, the inverse temperature of
the generated Gibbs state can reach negative values in principle (physically
corresponding to a state with an inverted population). The uncertainty
on the temperature of the Gibbs state is bounded by
\begin{equation}
\begin{array}{rcl}
\delta\beta & \leq & \frac{2^{2-q}}{\eta}\left(1+\frac{\left\Vert \mathcal{H}'\right\Vert _{\infty}}{\left\Vert \mathcal{H}_{B}\right\Vert _{\infty}}\right)\\
\\
 & = & 2^{2-q}\sqrt{\frac{\lambda}{m}}\frac{1}{\left\Vert \mathcal{H}'\right\Vert _{\infty}}\left(1+\frac{1}{\sqrt{m\lambda}}\right).
\end{array}\label{eq:InverseTemperatureUncertainty}
\end{equation}
At least $q$ qubits are needed according to the rule 
\begin{equation}
q\geq\left\lceil -\log_{2}\left(\frac{\delta\beta\eta}{1+\frac{\left\Vert \mathcal{H}'\right\Vert _{\infty}}{\left\Vert \mathcal{H}_{B}\right\Vert _{\infty}}}\right)+2\right\rceil \label{eq:RequiredAncillaTemperaturePrecision}
\end{equation}
 and the average number of runs required to achieve some inverse temperature
is

\begin{equation}
\overline{\sharp\textrm{runs}}\leq2^{q}\sqrt{\frac{\pi}{2m}}e^{\frac{2}{\lambda}+\beta\left\Vert \mathcal{H}'\right\Vert _{\infty}+\frac{\lambda\left\Vert \mathcal{H}'\right\Vert _{\infty}^{2}\beta^{2}}{8}}.\label{eq:ExpectedNumberOfRuns}
\end{equation}
This last bound is a worst-case scenario as finding the ground state
of the Fermi-Hubbard is in general a $\mathrm{QMA-hard}$ problem.


\end{document}